\documentclass[twoside]{article}
\usepackage[preprint]{aistats2027}
\usepackage[T1]{fontenc}
\usepackage[utf8]{inputenc}
\usepackage{amsmath,amssymb,booktabs,array,multirow,graphicx}
\usepackage[table]{xcolor}
\usepackage[round]{natbib}
\usepackage{lmodern}
\usepackage{microtype}
\usepackage{balance}
\usepackage{tabularx}
\usepackage[hidelinks]{hyperref}
\hypersetup{pdftitle={The Implementation Lottery: Auditing Idea Reliability in Automated Research},
 pdfauthor={Jingjie Ning, Shanshan Zhong, Xiaochuan Li, Ji Zeng, Chenyan Xiong}}
\definecolor{ideablue}{HTML}{5271AE}
\definecolor{seedblue}{HTML}{70ACDE}
\definecolor{audityellow}{HTML}{F5CC7D}
\definecolor{implorange}{HTML}{FFA660}
\definecolor{decisionred}{HTML}{D85B59}
\newcommand{\Q}{\mathcal{Q}}
\newcommand{\E}{\mathbb{E}}
\newcommand{\Var}{\operatorname{Var}}

\newcommand{\tabref}[1]{Table~\ref{#1}}
\newtheorem{proposition}{Proposition}
\newcommand{\Cov}{\operatorname{Cov}}
\begin{document}
\runningtitle{The Implementation Lottery}
\runningauthor{Jingjie Ning et al.}
\raggedbottom
\twocolumn[
\aistatstitle{The Implementation Lottery:\\
Auditing Idea Reliability in Automated Research}
\aistatsauthor{Jingjie Ning, Shanshan Zhong, Xiaochuan Li, Ji Zeng, Chenyan Xiong}
\aistatsaddress{School of Computer Science, Carnegie Mellon University\\
\texttt{\{jening,szhong2,xiaochu4,jizeng,cx\}@cs.cmu.edu}}
]
\begin{abstract}
Automated research agents use program scores to judge ideas. We call variation in this evidence across implementations the implementation lottery. We introduce an Idea Reliability Audit that freezes mechanism specifications, samples independent programs, and compares selected code with fresh implementations of its mechanism. Across 3,048 assignments on 31 tabular classification tasks, all four primary aggregation tests have Holm-adjusted $p\geq0.56$. Under mean-of-five selection, the prespecified secondary intention-to-treat comparison gives selected-code premiums of 0.38 [0.08, 0.82] and 0.45 [0.05, 1.12] accuracy-equivalent points for Bounded and Agentic execution, respectively. Minimum task-deletion means are 0.20 and 0.14. Fidelity conditioning exposes concentration: the Agentic premium falls from 0.33 to 0.01 when one task is removed. Post-outcome analysis finds cross-split variation on 41 of 70 paired cards per process. Exploratory replay gives nearly equal Bounded point losses at four and twenty implementations; Agentic point losses decrease across the four evaluated budget rules. Under duration costs, one seed per program minimizes fitted common-design variance. The two-seed design becomes preferable when implementation-to-seed cost ratios exceed approximately 13 or 9.5 in the continuous-budget model. The audit distinguishes evidence for reusing a selected artifact from evidence for implementing its idea again.
\end{abstract}

\section{Introduction}
An automated researcher proposes ideas, implements them, and continues the branch with the strongest experimental result. Systems for scientific discovery increasingly use this decision to direct exploration and allocate compute \citep{lu2024aiscientist,yamada2025aiscientistv2,lu2026automation,si2026executiongrounded}. Each observed result is produced by a concrete program. We call variation in idea-level evidence across implementations the \emph{implementation lottery}. Promoting the underlying idea requires evidence about other programs that could realize the same specification.

\begin{figure*}[!t]
\centering
\includegraphics[width=\textwidth]{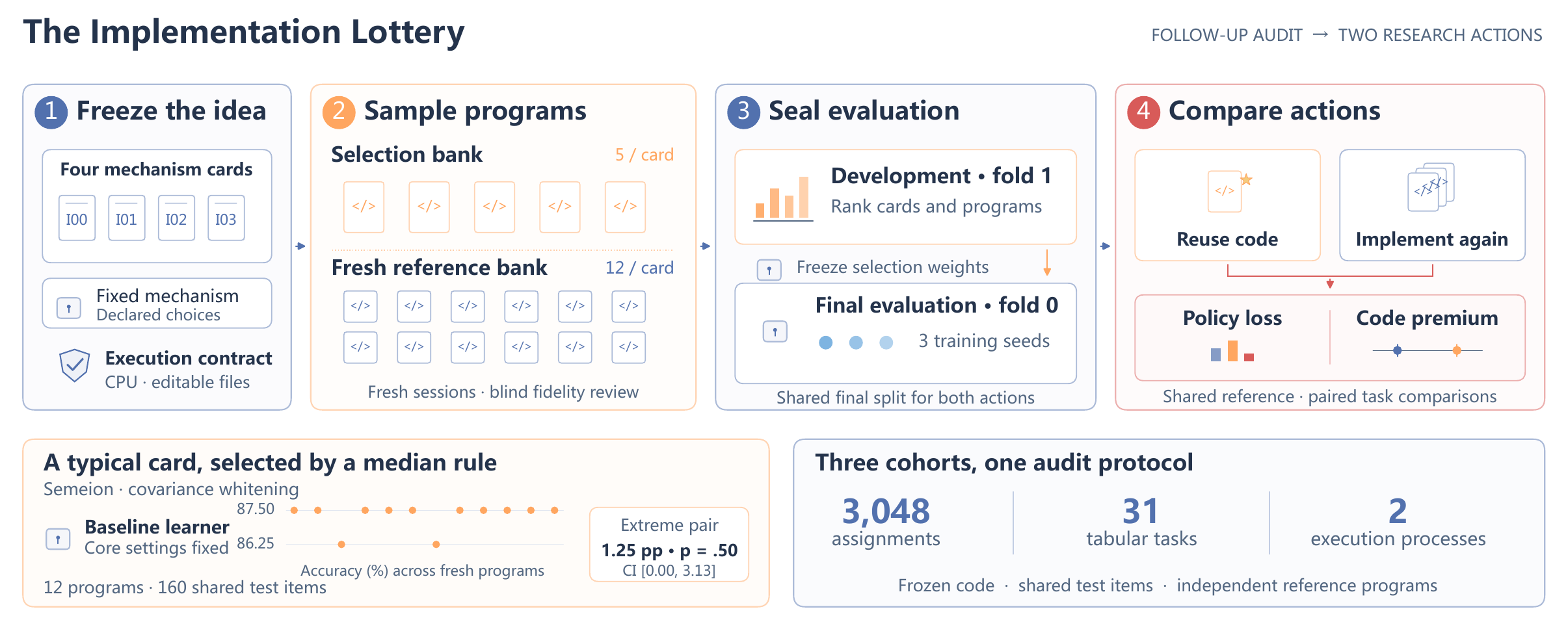}
\caption{Follow-up audit design and a median-rule example. Bounded semeion whitening is rank 34 of 68 eligible positive-SD reference cards. Its twelve model-judged faithful programs preserve all five checked gradient-boosting settings. The central pair has identical predictions; the extreme pair differs by 1.25 points [0.00, 3.13], with exact McNemar $p=0.50$ \citep{mcnemar1947paired}. Appendix~\ref{app:premium-robustness} gives the selection rule and paired counts.}
\label{fig:overview}
\end{figure*}

A mechanism specification leaves implementation choices open. Figure~\ref{fig:overview} shows the follow-up card at the median positive implementation standard deviation among complete model-judged faithful cards with baseline learner settings. Its twelve programs score 86.25--87.50\% on the same 160 test items. The two central programs have identical predictions. This typical card illustrates both the permitted implementation choices and the resolution of a small shared test sample.

Two actions follow from a strong result. Reusing its code preserves the selected realization. Promoting its mechanism calls for fresh implementation. These actions can have different expected outcomes even after evaluation on a separate data split. Their difference measures how much a selected program's gain overstates fresh-draw idea quality. Candidate gaps then determine whether implementation variation changes the preferred idea, and the chosen cost unit determines which form of replication is affordable.

Our contribution combines an audit protocol, evidence of transported implementation differences, and a direct comparison of artifact and idea selection. The protocol freezes cards and program choices, separates development selection from a shared final reference, and records both operational and model-fidelity-conditioned results. The prespecified secondary comparison measures selected-code optimism. Post-outcome diagnostics examine cross-split association and model calibration; exploratory cost analysis and focused expansion quantify replication tradeoffs. Classical statistics define the measured targets.

\section{Related Work}
\paragraph{Research agents and implementation.}
Automated research systems combine ideation, execution and reporting \citep{lu2024aiscientist,schmidgall2025agentlab,yuan2025dolphin}. Tree and execution-grounded search allocate experiments \citep{yamada2025aiscientistv2,si2026executiongrounded}. Co-Scientist refines hypotheses \citep{gottweis2026coscientist}, while specialist agents share measured trial histories to revise training recipes \citep{ning2026specialists}. MLGym, MLR-Bench and ResearchGym evaluate broader workflows \citep{nathani2025mlgym,chen2025mlrbench,garikaparthi2026researchgym}. Molecular and materials Auto Research test selected workflow changes on held-out data \citep{ning2026molecular,ning2026materials}. Ideation--execution studies evaluate proposed ideas \citep{si2025ideas,si2026gap}; ResearchCodeBench, LMR-BENCH and AutoExperiment examine implementation fidelity \citep{hua2025researchcodebench,yan2025lmrbench,kim2026autoexperiment}. Our audit samples independent programs for each frozen mechanism.

\paragraph{Variation and measurement.}
Multiverse and many-analyst research characterize analytic multiplicity \citep{steegen2016multiverse,silberzahn2018many,bell2022multiverse}. \citet{bertran2026many} measure variation among autonomous analysts given fixed data and a hypothesis. Underspecification exposes multiple predictors satisfying common requirements \citep{damour2022underspecification}. Benchmark studies separate sources of variation and task effects \citep{bouthillier2021variance,dehghani2021benchmarklottery}. Repeat-aware RAG audits measure corpus-induced answer churn against same-snapshot repeat variability \citep{ning2026answerchurn}. Repeated-measurement tools include agent reliability, intraclass correlation and generalizability theory \citep{rabanser2026reliability,mustahsan2025icc,shrout1979icc,brennan2001generalizability}. We apply them with explicit construct admission and model-based fidelity labels \citep{bean2025construct,zheng2023judge}.

\paragraph{Selection, evidence, and replication.}
Ranking-and-selection methods relate measurement noise to correct selection \citep{bechhofer1954selection,kim2001selection}. The optimizer's curse and model-selection overfitting characterize selection optimism \citep{smith2006optimizer,cawley2010overfitting}. Gaussian maximum formulas and multistage allocation are established results \citep{clark1961greatest,cochran1977sampling}. The Discovery Certification Protocol tests outcome recovery under matched information boundaries \citep{ning2026dcp}; WhatWorkedBench tests component-effect predictions after budgeted experiments \citep{ning2026whatworked}. Our audit measures the gap between selected artifacts and fresh realizations of a fixed mechanism. Program search demonstrates the value of artifact preservation \citep{romera2024funsearch,novikov2025alphaevolve}. Task-level intervals follow uncertainty-aware evaluation practice \citep{agarwal2021precipice,henderson2018matters}.

\section{Audit Design and Statistical Interpretation}
Classical ranking and maximum identities define the audit's comparison targets and replication rules. Appendix~\ref{app:proofs} derives these calculations.

\subsection{Idea quality and a shared reference}
A task $s$ fixes data, a baseline, editable files, and an evaluation contract. A mechanism card $h_i$ declares required behavior and open implementation choices. Executor $e$ samples program $m\sim P_e(m\mid s,h_i)$. Its operational idea quality is
\begin{equation}
\Q_e(s,h_i)=\E_{m,r}[U(s,h_i,m,r)],
\label{eq:quality}
\end{equation}
where $r$ denotes training randomness and $U$ includes the declared treatment of failures. Instructions, budget, and the implementation process define the target.

Development evidence $D$ fixes selection probabilities $\pi_i$ over a frozen candidate set. Fresh programs estimate final mean utilities $\mu_i$ with reference means $q_i$. Empirical-reference loss is
\begin{equation}
L(\pi)=\max_i q_i-\sum_i\pi_i q_i.
\label{eq:loss}
\end{equation}

\begin{proposition}[Shared-reference comparison]
\label{prop:reference}
Assume $\E[\mathbf q\mid D]=\boldsymbol\mu$ and finite covariance $V=\Cov(\mathbf q\mid D)$. For policies fixed by $D$,
\begin{align}
\E[L(\pi)\mid D]&=B_{\rm ref}+R(\pi),\label{eq:floor}\\
L(\pi)-L(\pi')&=\mathbf d^{\mathsf T}\mathbf q,\label{eq:contrast}\\
\E[\mathbf d^{\mathsf T}\mathbf q\mid D]&=\mathbf d^{\mathsf T}\boldsymbol\mu,\quad
\Var(\mathbf d^{\mathsf T}\mathbf q\mid D)=\mathbf d^{\mathsf T}V\mathbf d,\nonumber
\end{align}
where $\mathbf d=\pi'-\pi$, $R(\pi)=\max_i\mu_i-\pi^{\mathsf T}\boldsymbol\mu$, and $B_{\rm ref}=\E[\max_i q_i\mid D]-\max_i\mu_i\geq0$.
\end{proposition}
The maximum's noise floor cancels exactly from a shared-reference policy contrast. This property motivates our primary endpoint. Appendix~\ref{app:estimation} specifies the archived code-versus-seed diagnostic and \mbox{its two sampling targets}.

\subsection{Gaps, variance, and implementation budgets}
For mean-based selection, consider independent Gaussian program and seed effects,
\begin{equation}
Y^D_{ijk}=\mu_i^D+M^D_{ij}+R^D_{ijk},\qquad
v_i(J,K)=\frac{\sigma_{Mi}^2}{J}+\frac{\sigma_{Ri}^2}{JK}.
\label{eq:budget}
\end{equation}
Here $J$ is the number of implementations and $K$ the seeds per program. The variance identity holds for independent effects with finite second moments.

\begin{proposition}[Two-card selection]
\label{prop:gap}
Suppose card 1 has final mean advantage $\Delta_T>0$ and development advantage $\Delta_D$. For $s_D^2=v_1(J,K)+v_2(J,K)>0$,
\begin{equation}
\begin{aligned}
\Pr(\hat i=2)&=\Phi(-\Delta_D/s_D),\\
\E[R_T]&=\Delta_T\Phi(-\Delta_D/s_D).
\end{aligned}
\label{eq:gaprisk}
\end{equation}
\end{proposition}
The candidate gap and measurement noise jointly determine selection risk. Development-to-final transfer enters through the two gaps. Appendix~\ref{app:proofs} extends the bound to every competitor in a larger portfolio and derives the Gaussian finite-reference floor.

Maximum aggregation targets an upper order statistic. For independent $Z_{ij}\sim N(\mu_i,s_i^2)$,
\begin{equation}
\E[\max_{j\leq J}Z_{ij}]=\mu_i+s_i c_J,\qquad
c_J=\E[\max_{j\leq J}N_j(0,1)].
\label{eq:max}
\end{equation}
At equal means, larger variance raises the expected maximum for $J\geq2$. Mean and maximum policies therefore require separate empirical evaluation.

Replication also has a cost. With per-program cost $c_M$, per-seed cost $c_R$, and budget $B=J(c_M+Kc_R)$, minimizing the variance of a card-mean estimate gives
\begin{equation}
K^*=\frac{\sigma_R}{\sigma_M}\sqrt{\frac{c_M}{c_R}}.
\label{eq:allocation}
\end{equation}
The continuous-budget variance at $K=2$ is lower than at $K=1$ exactly when $c_M/c_R>2\sigma_M^2/\sigma_R^2$ for positive seed variance. This is the classical multistage allocation calculation \citep{cochran1977sampling}. We evaluate feasible integer $(J,K)$ pairs under measured costs. The rule allocates effort to estimating fresh-draw quality; a separate replay measures the decision consequences of concentrating implementations on close competitors.

\subsection{Selected programs as idea-level evidence}
Development scores fix program weights $w_{ij}$ and card weights $\pi_i=\sum_j w_{ij}$. With final program scores $T_{ij}$ and fresh card means $q_i$, the observed premium is
\begin{equation}
G=\sum_{ij}w_{ij}T_{ij}-\sum_i\pi_iq_i.
\label{eq:observed-premium}
\end{equation}
The premium compares the selected artifact's result with fresh-draw idea quality. Selection across cards can produce a premium from one draw per card.

For the five-program selection bank, adding and subtracting its card means $\bar T_i$ gives a within-bank advantage $\sum_{ij}w_{ij}(T_{ij}-\bar T_i)$ and a weighted bank gap $\sum_i\pi_i(\bar T_i-q_i)$. Equal-card bank comparisons measure a separate mean shift. Appendix~\ref{app:proofs} develops a Gaussian regression approximation; task-held-out checks evaluate its predictive accuracy.

\begin{table*}[!t]
\centering\small
\caption{Cohort design and implementation standard deviation in accuracy points. Intervals are 95\% task-bootstrap bounds. The archived population also requires construct admission; prospective populations require primary model-judged fidelity and complete seed evaluations.}
\label{tab:evidence}
\renewcommand{\arraystretch}{1.12}
\rowcolors{2}{white}{ideablue!3}
\begin{tabularx}{\textwidth}{@{}l>{\raggedright\arraybackslash}Xll@{}}
\toprule
\rowcolor{ideablue!8} \textbf{Cohort} & \textbf{Sampling design} & \textbf{Bounded SD} & \textbf{Agentic SD} \\
\midrule
Archived, 13 tasks & Three programs per card; post-implementation admission & 1.10 [0.00, 1.72] & 2.65 [0.00, 3.86] \\
Initial, 12 tasks & Prospective admission; five selection and three reference programs & 2.16 [0.38, 3.60] & 0.93 [0.31, 1.37] \\
Follow-up, 18 tasks & Development selection; twelve fresh programs per card & 1.17 [0.46, 1.70] & 0.98 [0.58, 1.40] \\
\bottomrule
\end{tabularx}
\end{table*}

\section{Experimental Design}
\subsection{Cohorts and execution processes}
The three cohorts use OpenML-CC18 tabular classification tasks \citep{vanschoren2014openml,bischl2021suites} and a scikit-learn histogram gradient-boosting baseline \citep{pedregosa2011sklearn}. Agents edit features and models under a single-predictor CPU policy. Fold 0 supplies final evaluation; labels stay inside the controller. \emph{Bounded} permits four calls, 64,000 branch tokens, six inspections, and two repairs. \emph{Agentic} uses a fresh Read/Edit SDK session with a \mbox{requested 20-turn option}. 

\begin{center}
\fcolorbox{ideablue!30}{ideablue!4}{\parbox{.94\columnwidth}{\small
\textbf{Audit terms.} A \emph{card} fixes a mechanism and permitted choices. Prospective \emph{admission} precedes implementation. \emph{Model-judged faithful} requires the primary blind judge's faithful label; \emph{both-review faithful} requires both labels. Executability is assessed separately. \emph{ITT} includes every assignment and its operational utility. \emph{KEEP} preserves the baseline.}}
\end{center}

The archived cohort contains 312 assignments on 13 tasks generated with DeepSeek V4 Pro. Post-implementation semantic screening admits 43 of 52 cards under construct-focused adjudication. The initial prospective cohort contains 768 assignments on twelve new eligible tasks, with eight programs per card and process. Five programs supply selection evidence and three form the reference, both evaluated on fold 0. Its four mechanism families are early stopping, feature filtering, covariance whitening, and an RBF classifier.

The follow-up adds six tasks and 1,968 assignments, bringing the total to 3,048 unique attempts on 31 tasks. Every task/process has five selection programs and twelve fresh reference programs per card. Continuing tasks reuse 480 initial selection programs. Selection trains on folds 2--9 and scores fold 1; final evaluation trains on folds 1--9 and scores fold 0. New code and model judgments freeze before evaluation, and policy weights freeze before opening new final outcomes. The twelve continuing tasks' earlier fold-0 outcomes had already been observed. All prospective model roles use \texttt{deepseek-flash}; two blind prompts assess fidelity, with empty-patch and mismatched-card controls.

The four primary tests and secondary selected-code comparison were specified locally before new final outcomes. The plan compares the highest development-score program with fresh implementations. Policy weights froze at 12:28:58 UTC on October 2; new final evaluation began at 12:30:11. Premium robustness, transport, calibration and allocation analyses were added after outcomes (Appendix~\ref{app:timeline}).

The follow-up was designed after the initial results. Its evidence covers one prospective model family, tabular CPU classification, four mechanism families, and one final fold. Fidelity judgments come from two LLM prompts. These choices define the scope of the observed implementation distributions and decisions.

\subsection{Utility, policies, and uncertainty}
Let $V=1$ indicate that execution produced a valid score. For baseline accuracy $a_s$, program accuracy $a$, and fidelity label $\ell$,
\begin{equation}
U=\begin{cases}
(a-a_s)/a_s-p(\ell),& V=1,\\
-0.10,& V=0.
\end{cases}
\label{eq:utility}
\end{equation}
The fidelity penalty is zero for faithful or partial labels, 0.02 for drift or missing labels, and 0.10 for a fidelity-invalid label. Thus an executable program with a fidelity-invalid label receives its measured gain minus 0.10. Invalid patches, runtime errors, timeouts, and missing assigned seeds receive failure utility. Partial attempts contribute their measured operational outcome; fidelity-conditioned analyses require faithful labels. Utility differences multiplied by $100a_s$ are accuracy-equivalent points. The saved banks also support alternative failure and fidelity penalties.

Five fixed policies use one program, the mean or maximum of three, and the mean or maximum of five. The follow-up enumerates every within-card subset and cross-card combination. The resulting distributions specify both a selected card and its highest-scoring observed program. Exact card ties favor the largest label; program ties receive uniform weights. Shared references average twelve assignments and three seeds, including failure outcomes.

Primary pooled ITT tests compare Max3 with Mean3 and Single with Mean5. Exact paired sign flips use Holm adjustment across both contrasts and both processes. Task-bootstrap intervals use 20,000 resamples with equal task weights. Enumeration worlds stay grouped within their task. The supplement gives task counts, attainable test resolution, leave-one-task-out ranges, and the two fidelity-conditioned populations.

The exploratory top-two focused-expansion replay starts with one program for each of four cards. Its gated variant uses a leading-gap threshold. If the leading gap is at most $\Phi^{-1}(0.95)$ times its estimated noise standard deviation, the two leading cards expand to five programs each. Selection then compares their means with the other cards' original scores. Each task's variance estimate uses development banks from the other seventeen tasks, separately by mechanism family. This rule uses four or twelve programs. Comparators include Mean5 with twenty, unconditional expansion of the top two with twelve, and random expansion at the same expected per-task budget. All choices are scored against the primary shared reference.

\begin{figure*}[!t]
\centering
\includegraphics[width=.95\textwidth]{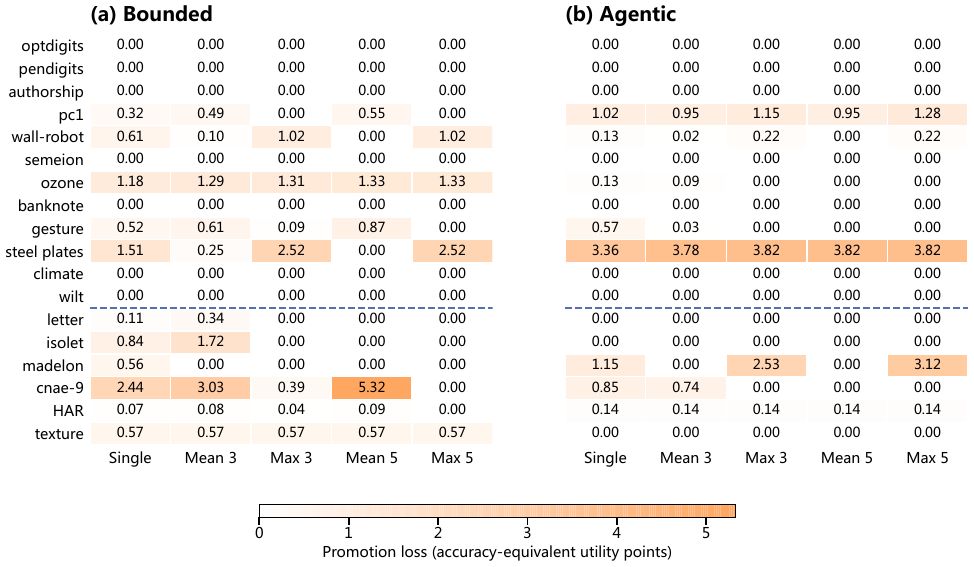}
\caption{Follow-up ITT empirical-reference loss in accuracy-equivalent points. The dashed line separates twelve continuing tasks from six new tasks. Seven Bounded and ten Agentic tasks have zero loss under every policy. Colors share one scale, and cell values show the complete task-level matrix. Shared-reference policy contrasts use paired differences between columns.}
\label{fig:extension-task}
\end{figure*}

\section{Results}
\subsection{Primary policy comparisons}
The four primary pooled ITT tests have Holm-adjusted $p\geq0.5625$. Max3-minus-Mean3 loss is $-0.14$ [$-0.59$, 0.29] for Bounded and 0.12 [$-0.10$, 0.44] for Agentic. Single-minus-Mean5 is 0.00 [$-0.42$, 0.34] and 0.14 [$-0.01$, 0.31]. Their positive/tied/negative task counts are 3/9/6, 4/11/3, 5/8/5, and 6/11/1. Figure~\ref{fig:extension-task} shows every task and policy.

CNAE-9 contributes 61\% of Bounded Mean5 loss. Deleting one task moves its full mean of 0.485 through [0.200, 0.514]; the Agentic range is [0.064, 0.289] around 0.273. The initial cohort has zero to four nonzero task differences per ITT contrast. Exact sign-flip resolution and task-bootstrap intervals expose this concentration directly. Baseline-headroom and failure-penalty analyses give further operational views. Changing failure utility from 0 to $-0.20$ moves the frozen-policy Bounded Single-minus-Mean5 difference from $-0.065$ to 0.065 points.

Baseline KEEP supplies another action. Under Mean5, it reduces negative-reference promotions from 33.3\% to 16.7\% and from 22.2\% to 11.1\%, while declining a positive-reference candidate on 27.8\% and 16.7\% of tasks. A separately recorded runtime-load diagnostic rescues 29 of isolet's 55 primary timeouts and changes pooled seed-zero policy losses by at most 0.018 points. These results, all five fixed policies, and the original primary outcomes appear in Appendix~\ref{app:followup-results}.

\subsection{Prespecified secondary selected-code comparison}
The retained-versus-fresh endpoint and highest-score retention rule were specified before new outcomes, with program weights frozen after development evaluation. Under Mean5, selected-code gains average 0.48 and 0.80 points above baseline; fresh realizations of the selected cards average 0.10 and 0.35. The ITT premiums are 0.38 [0.08, 0.82] and 0.45 [0.05, 1.12]. Single gives 0.27 [0.11, 0.45] and 0.21 [0.01, 0.53]. Table~\ref{tab:premium-main} places both workflows beside their fidelity-conditioned estimates.

The added robustness audit finds positive ITT Mean5 means after every task deletion: 0.20--0.40 for Bounded and 0.14--0.48 for Agentic. Across all deleted samples, the smallest bootstrap lower endpoints are 0.049 and 0.030 points. The task medians are only 0.031 and 0.024, showing that the mean premium exceeds the typical task's value. Exact sign-flip probabilities are 0.0068 and 0.0044. These diagnostic tests use a separate, unadjusted family from the four primary tests.

\begin{figure*}[!t]
\centering
\includegraphics[width=\textwidth]{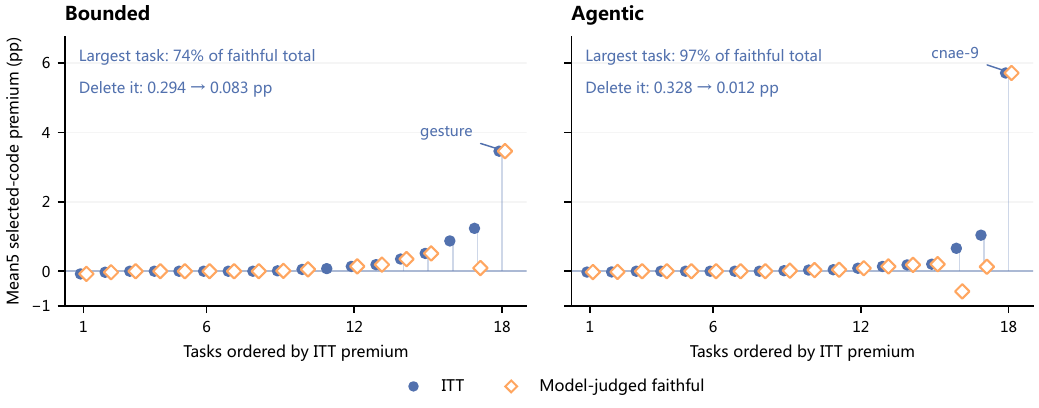}
\caption{Task concentration of the prespecified Mean5 premium. Each position is one task, ordered by its ITT premium; hollow diamonds give the model-judged faithful result where the complete-bank population is available. The largest task contributes 74\% and 97\% of the signed faithful total. Callouts show the faithful mean before and after deleting that task. All eighteen ITT tasks appear in each panel; Bounded has sixteen faithful tasks.}
\label{fig:premium-concentration}
\end{figure*}

Figure~\ref{fig:premium-concentration} shows how fidelity conditioning changes task concentration. Bounded's largest task contributes 74\% of the signed Mean5 total; Agentic's contributes 97\%. Deleting the latter reduces its mean from 0.328 to 0.012 points. Agentic Single has exact $p=0.080$ in ITT and $p=0.414$ after fidelity conditioning.

\begin{table*}[!ht]
\centering\small
\renewcommand{\arraystretch}{1.16}
\rowcolors{2}{white}{ideablue!3}
\caption{Prespecified secondary premiums with post-outcome robustness checks. P/T/N counts positive, tied, and negative task premiums. Deletion ranges contain task-deletion point means. Faithful means the complete-bank primary model-judged population (16 Bounded and 18 Agentic tasks); ITT contains 18 each. Exact sign-flip probabilities are unadjusted diagnostic values. All score differences are accuracy-equivalent points.}
\label{tab:premium-main}
\begin{tabularx}{\textwidth}{@{}>{\raggedright\arraybackslash}X>{\raggedright\arraybackslash}X>{\raggedright\arraybackslash}Xrrclr@{}}
\toprule
\rowcolor{ideablue!8} \textbf{Process} & \textbf{Population} & \textbf{Policy} & \textbf{Mean [95\% CI]} & \textbf{Median} & \textbf{P/T/N} & \textbf{Deletion means} & \textbf{Raw $p$} \\
\midrule
Bounded & ITT & Single & 0.268 [0.109, 0.453] & 0.052 & 11/4/3 & [0.216, 0.294] & 0.005 \\
Bounded & ITT & Mean5 & 0.377 [0.082, 0.820] & 0.031 & 10/6/2 & [0.196, 0.404] & 0.007 \\
Bounded & Faithful & Single & 0.096 [0.022, 0.184] & 0.026 & 8/5/3 & [0.067, 0.108] & 0.035 \\
Bounded & Faithful & Mean5 & 0.294 [0.030, 0.749] & 0.005 & 8/6/2 & [0.083, 0.318] & 0.025 \\
\addlinespace[2pt]
Agentic & ITT & Single & 0.215 [0.010, 0.532] & 0.000 & 8/6/4 & [0.078, 0.247] & 0.080 \\
Agentic & ITT & Mean5 & 0.448 [0.050, 1.124] & 0.024 & 10/6/2 & [0.138, 0.476] & 0.004 \\
Agentic & Faithful & Single & 0.156 [-0.079, 0.489] & 0.000 & 7/6/5 & [0.015, 0.207] & 0.414 \\
Agentic & Faithful & Mean5 & 0.328 [-0.049, 0.996] & 0.007 & 9/6/3 & [0.012, 0.382] & 0.431 \\
\bottomrule
\end{tabularx}
\end{table*}

\begin{figure*}[!t]
\centering
\includegraphics[width=\textwidth]{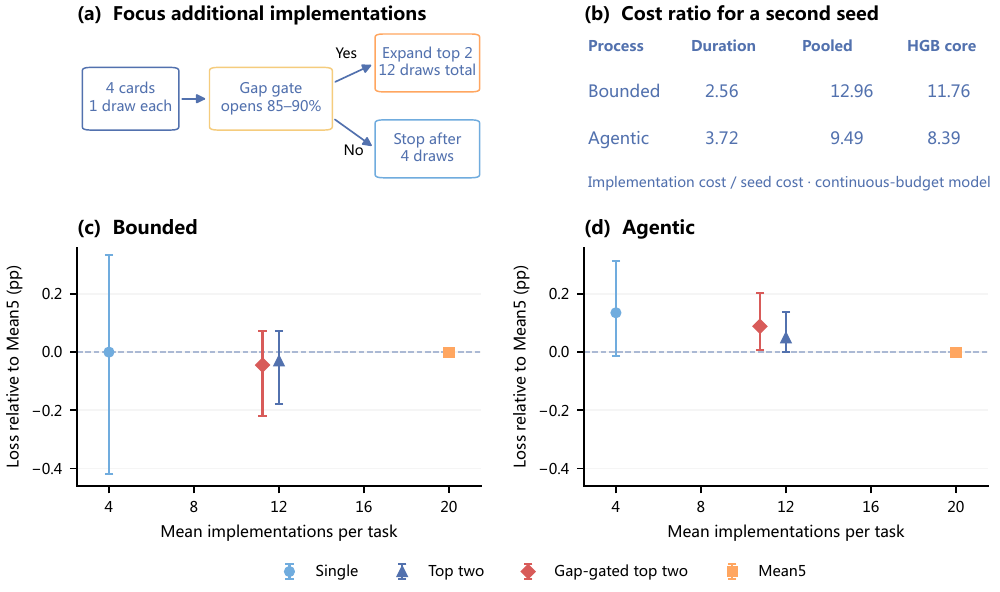}
\caption{Exploratory replication choices. The upper panels show focused expansion and the implementation-to-seed cost ratios that switch the continuous-budget choice from one seed to two. The lower panels compare implementation count with paired loss relative to Mean5. Positive values mean higher loss. The gap-gated rule uses a nominal 5\% Gaussian threshold and task-held-out development variances. Intervals resample tasks; Table~\ref{tab:allocation-replay} gives exact tests and the budget-matched expansion comparator.}
\label{fig:allocation}
\end{figure*}

Starting from the reported development gain adds another measurable component. For Mean5, selected development minus selected final gain is 0.24 and 0.68 points; adding the final retained-minus-fresh premium gives total development-to-fresh gaps of 0.61 and 1.12. The first term includes split transfer and a change in training-set size. Appendix~\ref{app:premium-robustness} reports both terms for all policies and populations.

\subsection{Post-outcome transport and bank diagnostics}
Model-judged faithful programs show implementation standard deviations around one accuracy point in the follow-up (Table~\ref{tab:evidence}). Across cohorts and processes, 69--83\% of complete programs have equal accuracy at all three assigned seeds.

Within-card deviations correlate across splits at 0.59 [0.44, 0.73] and 0.68 [0.36, 0.83]. We average card covariance matrices within tasks, then across tasks; higher-variance cards have more influence on the resulting correlation. Task-deletion ranges are [0.55, 0.63] and [0.57, 0.74]. Each process has 41 of 70 paired cards (59\%) that vary on both splits; 29 are flat on at least one split. Informative cards have median Spearman correlations of 0.75 and 0.79. Agentic continuing tasks have median rank correlation 0.23 and trimmed covariance correlation 0.48. Removing deviations over ten accuracy points from the card median gives pooled correlations of 0.59 in both processes.

Static source analysis tags all 3,048 assignment records by five core gradient-boosting settings. In follow-up reference programs with those settings fixed at baseline, implementation standard deviations are 1.28 [0.41, 1.89] and 1.04 [0.55, 1.54] points. Thus variation also appears through the specified feature and stopping choices. Paired item analysis provides complementary evidence. Subtracting the estimated common-test-sample component leaves mean across-program dispersion of 1.23 [0.18, 2.67] and 0.79 [0.26, 1.58] squared points. The source strata, paired-item assumptions, and program coverage appear in the supplement.

A separate convergence diagnostic reruns eight source-identified SVC programs and reproduces every saved prediction vector. All seven literal 300-iteration caps generate convergence warnings. The gesture program, with 169/988 correct predictions, reaches its cap in all ten binary subproblems. The eighth program resolves its dynamic cap to $-1$ and converges. These fits document numerical nonconvergence within the sampled implementation distribution (Table~\ref{tab:svc-convergence}).

Equal-card continuing-bank minus fresh differences are $-0.02$ [$-0.20$, 0.13] and $-0.05$ [$-0.13$, 0.03]. The new tasks give $-0.42$ [$-1.04$, 0.10] and $-0.02$ [$-0.42$, 0.33]. Thus the full selection banks have no observed positive mean shift. The selected-card Bounded continuing gap is 0.11, reflecting different weights. Batch effects on selected programs remain mixed with finite draws and operational failures. Both banks share final folds and scoring contracts; continuing banks were evaluated in separate collection campaigns.

\subsection{Post-outcome model calibration}
We average reversal rates over all $\binom{12}{6}=924$ fit/held-out program partitions and every held-out subset choice. Both deterministic label orders and disjoint winner sets expose the effect of ties. Mean5 reversal rates are 0.174 and 0.123 under the original largest-label rule, compared with Gaussian complete-procedure predictions of 0.128 and 0.076. Reversing the tie rule gives 0.128 and 0.124; disjoint winner-set rates are 0.122 and 0.111. Thus Bounded calibration depends strongly on ties, while Agentic shows a discrepancy under all three definitions. Every partition reuses the same twelve programs. Model discrepancies can reflect score distributions, estimated components and dependence.

For selected-code premiums, each mechanism family uses a pooled slope fitted on the other tasks. Mean5 predictions are 0.11 and 0.39 against faithful premiums of 0.29 and 0.33, with mean absolute task errors of 0.28 and 0.12. These errors limit the regression formula's usefulness as a task-level predictor. Appendix~\ref{app:decision-diagnostics} reports the full policy and centering checks.

\subsection{Exploratory replication tradeoffs}
Recorded generation plus two-judgment durations average 35.25 and 45.61 seconds per reference program; seed evaluation averages 13.79 and 12.27 seconds. The resulting cost ratios, 2.6 and 3.7, lie below the pooled $K=1$ to $K=2$ switch thresholds of 13.0 and 9.5. Under these duration costs, one seed minimizes fitted variance at the common reference budget. Changing the common cost unit can change that choice. The baseline-HGB-core stratum has thresholds 11.8 and 8.4; every RBF reference program has zero observed three-seed score variation. Early-stopping cards instead have a seed-dominated fitted component (Table~\ref{tab:cost-switches}).

Top-two focused expansion spends twelve implementations; its gap-gated variant averages 11.2 and 10.8. The gate opens in 90.4\% and 84.7\% of worlds, so focusing on two cards supplies most of the saving from twenty. Bounded Single and Mean5 both have rounded mean loss 0.485, with paired difference interval [$-0.42$, 0.34]. These data give similar point losses and broad uncertainty about their difference.

Agentic mean losses decrease across the four evaluated rules: 0.408 with Single (four implementations), 0.361 with gated expansion (10.8), 0.322 with unconditional top-two expansion (twelve), and 0.273 with Mean5 (twenty). Gated expansion has higher loss than Mean5 by 0.09 [0.01, 0.20]; unconditional expansion differs by 0.05 [0.00, 0.14]. Figure~\ref{fig:allocation} displays paired uncertainty around this observed tradeoff.

\balance
\section{Implications for Automated Research}
The unit of evidence should match the next research action. An artifact result supports reuse of its code. Idea-level promotion calls for outcomes from fresh realizations under the declared executor and card. Reporting both exposes the selected-code premium and makes implementation replication part of the evidence for an idea ranking.

The audit also gives replication budgets an empirical basis. Absolute program and seed variance determine how replication changes precision, recorded costs determine feasible designs, and shared-reference replay measures the resulting selection tradeoff. Research-agent benchmarks and branch-promotion systems can use these quantities at the point where a program score becomes evidence for an idea.

\section{Conclusion}
The Idea Reliability Audit makes specification, program realization, and candidate promotion separately measurable. The primary comparisons provide no clear evidence of an aggregation advantage on this task set. Prespecified secondary comparisons show positive ITT selected-code premiums, with strong task concentration in the faithful Agentic population. Transport diagnostics and focused replay expose where implementation differences matter; cost-ratio thresholds make the assumptions behind replication choices explicit. Together, the protocol and measurements support evaluations of idea quality that reflect how the next implementation will be produced.

\clearpage
\nobalance
\twocolumn
\raggedbottom
\section*{AI Use Statement}
AI tools assisted with software development, experiment orchestration, analysis, literature verification, manuscript drafting, and figure production. The evaluated agents, model fidelity judges, and model-assisted source checks are documented with their roles, prompts, and model identifiers.
\bibliographystyle{apalike}
\bibliography{references}
\clearpage
\section*{Checklist}
\begin{enumerate}

  \item For all models and algorithms presented, check if you include:
  \begin{enumerate}
    \item A clear description of the mathematical setting, assumptions, algorithm, and/or model. [Yes]
    \item An analysis of the properties and complexity (time, space, sample size) of any algorithm. [Yes, Propositions 1--2 and the finite enumeration counts in Appendix B.]
    \item (Optional) Anonymized source code, with specification of all dependencies, including external libraries. [Yes]
  \end{enumerate}

  \item For any theoretical claim, check if you include:
  \begin{enumerate}
    \item Statements of the full set of assumptions of all theoretical results. [Yes]
    \item Complete proofs of all theoretical results. [Yes]
    \item Clear explanations of any assumptions. [Yes]
  \end{enumerate}

  \item For all figures and tables that present empirical results, check if you include:
  \begin{enumerate}
    \item The code, data, and instructions needed to reproduce the main experimental results (either in the supplemental material or as a URL). [Yes]
    \item All the training details (e.g., data splits, hyperparameters, how they were chosen). [Yes]
    \item A clear definition of the specific measure or statistics and error bars (e.g., with respect to the random seed after running experiments multiple times). [Yes]
    \item A description of the computing infrastructure used. (e.g., type of GPUs, internal cluster, or cloud provider). [Yes]
  \end{enumerate}

  \item If you are using existing assets (e.g., code, data, models) or curating/releasing new assets, check if you include:
  \begin{enumerate}
    \item Citations of the creator If your work uses existing assets. [Yes]
    \item The license information of the assets, if applicable. [Yes]
    \item New assets either in the supplemental material or as a URL, if applicable. [Yes]
    \item Information about consent from data providers/curators. [Not Applicable]
    \item Discussion of sensible content if applicable, e.g., personally identifiable information or offensive content. [Not Applicable]
  \end{enumerate}

  \item If you used crowdsourcing or conducted research with human subjects, check if you include:
  \begin{enumerate}
    \item The full text of instructions given to participants and screenshots. [Not Applicable]
    \item Descriptions of potential participant risks, with links to Institutional Review Board (IRB) approvals if applicable. [Not Applicable]
    \item The estimated hourly wage paid to participants and the total amount spent on participant compensation. [Not Applicable]
  \end{enumerate}

\end{enumerate}

\clearpage
\appendix
\onecolumn
\section{Decision Model and Proofs}
\label{app:proofs}
The results below specialize standard ranking, maximum, and regression identities to the audit's sampling targets. Their role is to state what each comparison identifies and to connect implementation budgets to decision risk. All expectations condition on the frozen task, evaluation contract, and executor unless stated otherwise. The mathematical Gaussian model supplies an approximation for successful numeric scores. Operational utilities with failure masses are handled directly by the empirical ITT analysis.

\subsection{Finite-reference bias and common-reference contrasts}
Fix a finite candidate set and let $D$ contain every observation used to form the policy weights. Assume the reference vector has finite second moments and satisfies $\E[\mathbf q\mid D]=\boldsymbol\mu$. This condition is met by independent reference draws from the target collection process; it also permits dependence across cards. With $L(\pi)=\max_i q_i-\pi^{\mathsf T}\mathbf q$, linearity yields
\[
\E[L(\pi)\mid D]=\E[\max_i q_i\mid D]-\pi^{\mathsf T}\boldsymbol\mu
=\underbrace{\E[\max_i q_i\mid D]-\max_i\mu_i}_{B_{\rm ref}}
+\underbrace{\max_i\mu_i-\pi^{\mathsf T}\boldsymbol\mu}_{R(\pi)}.
\]
The maximum is convex, so Jensen's inequality gives $B_{\rm ref}\geq0$. For another frozen policy $\pi'$, subtracting the two observed losses cancels the common maximum exactly. The expectation and variance of the resulting linear form are $\mathbf d^{\mathsf T}\boldsymbol\mu$ and $\mathbf d^{\mathsf T}V\mathbf d$, with $\mathbf d=\pi'-\pi$ and $V=\Cov(\mathbf q\mid D)$. These steps prove Proposition~\ref{prop:reference}. Scaling by $100a_s$ gives the paper's accuracy-equivalent unit, with the squared factor entering the variance.

Equality $B_{\rm ref}=0$ can occur when the same card attains the reference maximum almost surely. Otherwise the empirical optimum contributes a positive floor. Two loss measurements with different reference vectors generally have different floors. The saved-code target conditions on a chosen program and averages its training seeds. The fresh-code target averages program and seed draws. Equal card support and reference counts define a paired comparison of these two sampling targets.

Fidelity filters depend on generated programs and can alter candidate support. Conditional analyses describe those retained portfolios. Applying Proposition~\ref{prop:reference} to such a population requires a corresponding conditional-unbiasedness assumption; the fixed-portfolio ITT comparison supplies our primary endpoint. If the provider changes between collection batches, the empirical fresh-reference comparison targets the later collection process. Recorded timestamps and model aliases identify the sampled collection process.

\subsection{Gap-to-noise risk, split transfer, and reference floors}
In Equation~\ref{eq:budget}, average $J$ independent program effects and $JK$ independent seed errors. Variances add to $v_i(J,K)$. For two cards, $\bar Y_1^D-\bar Y_2^D$ is Gaussian with mean $\Delta_D$ and variance $s_D^2$. Selecting card 2 therefore has probability $\Phi(-\Delta_D/s_D)$. Its final-evaluation regret is $\Delta_T$, proving Equation~\ref{eq:gaprisk}. With $s_D=0$, selection follows the deterministic development means and the declared tie rule.

For $C>2$ cards with common development and final means, label a best card 1 and set $\Delta_i=\mu_1-\mu_i\geq0$. Independent Gaussian estimates give
\[
\E[R]=\sum_{i=2}^C\Delta_i\Pr(\hat i=i)
\leq\sum_{i=2}^C\Delta_i\Pr(\bar Y_i\geq\bar Y_1)
=\sum_{i=2}^C\Delta_i\Phi\!\left(-\frac{\Delta_i}{\sqrt{v_1+v_i}}\right).
\]
The bound uses every competitor. Zero-variance differences are evaluated by their deterministic limits, and zero gaps contribute zero regret. With different development and final means, each term instead uses the final gap as its multiplier and the signed development gap inside $\Phi$, assuming card 1 is final-optimal. Large variance in a low-ranked card can affect selection even when the top-two mean gap is large.

An exact target-level transfer decomposition further clarifies the interpretation. Let $i_D$ maximize development means and $i_T$ maximize final means. For any development-selected policy,
\[
R_T(\pi)=
\underbrace{\mu_{i_T}^T-\mu_{i_D}^T}_{\text{transfer cost of the development-optimal card}}
+\underbrace{\mu_{i_D}^T-\sum_i\pi_i\mu_i^T}_{\text{selection departure on the final target}}.
\]
The first term is nonnegative; the second can be negative because departing from the development-optimal card can improve final utility. Adding $B_{\rm ref}$ gives expected empirical-reference loss. The data do not identify these latent components separately without estimation assumptions. Matched score-bank sensitivities expose their combined effect on observed choices.

For jointly Gaussian reference means with card 1 best in expectation, the finite-reference floor is
\begin{equation}
B_{\rm ref}=s_q\phi(\Delta_T/s_q)-\Delta_T\Phi(-\Delta_T/s_q),\qquad s_q^2=\Var(q_1-q_2).
\label{eq:gaussianfloor}
\end{equation}
For the Gaussian reference formula, write $X=q_2-q_1\sim N(-\Delta_T,s_q^2)$. Since $\max(q_1,q_2)=q_1+X_+$,
\[
B_{\rm ref}=\E[X_+]
=\int_0^\infty x\frac{1}{s_q}\phi\!\left(\frac{x+\Delta_T}{s_q}\right)dx
=s_q\phi(\Delta_T/s_q)-\Delta_T\Phi(-\Delta_T/s_q).
\]
The formula permits correlated reference means through $s_q^2=V_{11}+V_{22}-2V_{12}$ and agrees with classical Gaussian maximum results \citep{clark1961greatest}. At $s_q=0$ the floor is zero. A zero gap gives positive reference loss but zero latent regret. The model plot displays this analytic floor for a perfect selector.

For maximum aggregation, write $Z_{ij}=\mu_i+s_iN_{ij}$ with independent standard normals. Monotonicity of a positive affine transformation gives Equation~\ref{eq:max}. For $J=2$, $\E[\max(N_1,N_2)]=\E[|N_1-N_2|]/2>0$; the expected maximum is nondecreasing as draws are added. Thus $c_J>0$ for $J\geq2$. This is a statement about the expected score at equal means. Selection probabilities also depend on competitor means and variances throughout the portfolio.

\subsection{Retained programs and selection intensity}
Assume independent paired program scores across cards and realizations, finite nonzero development variance, and a bivariate Gaussian distribution within each card. Their conditional regression is
\[
\E[T_{ij}\mid D]=\mu_i^T+\beta_i(D_{ij}-\mu_i^D),\qquad
\beta_i=\frac{\Cov(D_{ij},T_{ij})}{\Var(D_{ij})}.
\]
For an independent fresh bank from the same process, this gives
\begin{equation}
\E[G]=\E\!\left[\sum_{ij}w_{ij}(D)\beta_i(D_{ij}-\mu_i^D)\right].
\label{eq:retention}
\end{equation}
Let $w_{ij}(D)\geq0$ sum to one and define $\pi_i=\sum_jw_{ij}$. A fresh bank independent of $D$ has conditional expected selected-card utility $\sum_i\pi_i\mu_i^T$. Taking the conditional expectation of the weighted retained scores and subtracting this value proves Equation~\ref{eq:retention}. A linear conditional-mean assumption also suffices for this identity.

Single-draw policies can select program effects across cards. For $C$ equal-mean cards with common development standard deviation $\sigma_D$ and common slope $\beta$, selecting the best of one program per card gives expected gain $\beta\sigma_Dc_C=\rho\sigma_Tc_C$. This special case explains why within-card replication is unnecessary for a retained-code advantage. Unequal means or slopes require the full weighted expression.

For a fixed bank of $J=5$ programs per card, let $\bar T_i=J^{-1}\sum_jT_{ij}$. The empirical comparison has the exact decomposition
\[
G=\underbrace{\sum_{ij}w_{ij}(T_{ij}-\bar T_i)}_{A\text{, within-bank advantage}}
+\underbrace{\sum_i\pi_i(\bar T_i-q_i)}_{H\text{, bank-to-reference gap}}.
\]
The complementary nonretained bank uses weights $(\pi_i-w_{ij})/(J-1)$. Its gap is $H-A/(J-1)$. The bank difference combines finite draws and collection-batch effects. Cross-split covariance measures association between realized program outcomes, including their different training sets and test samples. A latent persistent-variance component would require a richer sampling model.

\subsection{Sampling uncertainty and finite test sets}
All task-level intervals resample complete tasks with replacement. The archived shared handwriting views also receive a source-cluster sensitivity. The tasks arise from stated metadata envelopes, so inference describes those selected task populations. Exact sign-flip tests enumerate all $2^m$ signs of nonzero paired task differences and require sign exchangeability under the null. For a two-sided absolute-mean statistic, at least two sign configurations are as extreme as the largest observed value, yielding minimum attainable $p=2^{1-m}$ for $m\geq1$. Ties can increase that minimum. With $m=0$, $p=1$.

Percentile intervals summarize the empirical task distribution. An all-zero sample generates only zero bootstrap means. For bias-corrected and accelerated intervals \citep{efron1987bootstrap}, the acceleration denominator is zero when all jackknife estimates agree. We report percentile intervals together with nonzero-task counts, attainable test resolution, and task-deletion ranges. A boundary interval and a discrete sign-flip test can therefore convey different information in a small subgroup.

For paired item uncertainty, fix the realized seed-zero programs and let $X_n\in\{0,1\}^P$ record their correctness on test item $n$. The supplement stores counts of each joint correctness pattern, preserving dependence across programs without benchmark labels or prediction arrays. A multinomial bootstrap over these pattern counts is equivalent to resampling the shared test items. It assumes independent identically distributed test items and conditions on tasks, fitted programs, policy weights, and failure/fidelity constants. Its intervals address a different uncertainty source from task resampling and from the primary three-seed reference.

Let $S_X$ be the unbiased sample covariance of the item correctness vectors, $N$ the number of test items, and $H=I-\mathbf1\mathbf1^{\mathsf T}/P$. Across-program accuracy variance is $\bar X^{\mathsf T}H\bar X/(P-1)$. Its estimated item-sampling contribution is $\operatorname{tr}(HS_X)/(N(P-1))$, since $\Cov(\bar X)=\Cov(X_n)/N$. We subtract this component from the observed dispersion and retain signed estimates. Cards are averaged within tasks, then tasks equally, with task-bootstrap intervals. The estimate concerns fixed realized programs under the item-sampling assumption. It includes their seed realizations and does not separately identify pure implementation variance. Multiplication by $10^4$ converts these variance estimates to squared accuracy points.

\section{Estimation and Decision Definitions}
\label{app:estimation}
The primary resampling unit is the task. Between-card effects use symbol $A$ throughout; I00--I03 are card identifiers. Within a task $s$, let $n_{si}$ be the number of retained implementations of card $i$, let $N_s=\sum_i n_{si}$, and let $C_s$ be the number of retained cards. Define the within-card sum of squares $W_s=\sum_{ij}(y_{sij}-\bar y_{si})^2$ and the between-card sum of squares $B_s=\sum_i n_{si}(\bar y_{si}-\bar y_s)^2$. Under independent card effects of variance $a$ and within-card effects of variance $b$,
\begin{align}
\E[W_s]&=b(N_s-C_s),\\
\E[B_s]&=b(C_s-1)+a\left(N_s-\frac{\sum_i n_{si}^2}{N_s}\right).
\end{align}
The second identity follows by expanding the weighted variance about the task mean. The within-card noise contributes $b(C_s-1)$ after centering; the shared effect within card $i$ contributes $a(n_{si}-n_{si}^2/N_s)$. Summing these contributions gives the expression above. Pooling sufficient statistics yields
\begin{align}
\hat b&=\frac{\sum_s W_s}{\sum_s(N_s-C_s)},\\
\hat a&=\frac{\sum_s B_s-\hat b\sum_s(C_s-1)}{\sum_s\left(N_s-\sum_i n_{si}^2/N_s\right)}.
\end{align}
We estimate seed variance from the mean sample variance of the available complete three-seed evaluations of saved programs. Implementation variance is $\hat b-\hat\sigma_R^2$. The reported components bound $\hat a$ and implementation variance below at zero; the analysis files include the raw values. Tasks with fewer than two retained cards supply no within-task between-card comparison. On a balanced design, these expressions equal the prespecified nested method-of-moments estimator. Unequal-cell estimates describe the retained population under the stated moment model; filtering can change its card and implementation distribution.

For $c$ complete cards with three implementations each, there are $3^c$ choices of one implementation per card. Each choice supplies a single selection decision and a reference formed from the other two implementations within each card. Relative permutation analysis fixes the first card's ordering and enumerates $(3!)^{c-1}$ alignments of the remaining cards. Each alignment averages its three matched-index decisions. Averaging those alignment means equals the full cross-card enumeration. Numerical verification confirms invariance to replicate relabeling.

The fixed tie rule selects the lexicographically largest card among exact maxima. A reference winner lies within $10^{-12}$ of the maximum reference score. The winner-set analysis uses this tolerance on both sides and records whether their intersection is empty. A sensitivity also averages disagreement and loss uniformly over the tied selection winners. Code and data include the selection and reference top-two margins, tied-decision fractions, loss quantiles, and the complete threshold grid.

For the saved-code control, each enumerated world first fixes one code realization per card. Each of its three seeds supplies the selection score once, with the other two seeds forming the reference. The implementation comparison and seed comparison therefore have the same card population and two reference draws per candidate. Their task means are paired before uncertainty estimation. The excess combines the two reference distributions' noise floors and program-conditional targets. The prospective version uses the first three frozen implementation slots for this matched diagnostic, while all eight slots enter variance estimation and independent-reference policy evaluation.

\subsection{Independent-reference follow-up}
The follow-up was designed after inspecting the initial prospective results. It reuses the exact accepted cards and first five code realizations on those twelve tasks, and adds six tasks selected by metadata. Each continuing task receives twelve new reference programs per card and process. Each new task receives five selection programs and twelve reference programs. This gives 1,968 new assignments, with 480 existing programs reused for selection. The six new task identifiers are 6, 3481, 9976, 9981, 14970, and 125922. The metadata envelope permits at most 20,000 rows, 900 numeric predictors, 26 classes, class imbalance 20, and complete predictors. Exposure exclusions and the exclusion of all multiple-features views are preserved.

Development selection trains each saved program on official folds 2--9 and evaluates it on fold 1. Final evaluation trains on folds 1--9 and evaluates on fold 0. All new code, reviews, and seed assignments are frozen before any new program receives either evaluation. Baselines receive the matching development evaluation. Classification accuracy is represented on the exact correct-count grid. Stored normalized test gains are converted back to that grid using the known test size and frozen baseline. The supplement preserves the original stored values alongside the analysis inputs.

For a policy using $k\in\{1,3,5\}$ implementations per card, each card contributes every size-$k$ subset of its five-program bank. The Cartesian product of these subsets defines selection worlds. With four cards, their counts are 625, 10,000, and one for $k=1,3,5$. A mean or maximum aggregation within each world, followed by the fixed card tie rule, defines selection probabilities $\pi_{si}^{(p)}$. All policy distributions and all card-subset sensitivities are saved before the reference outcomes are opened. Numerical ties use tolerance $10^{-12}$; the largest card label wins. Combinations share observations, and uncertainty resamples complete tasks.

Each reference program is assigned seeds 0, 1001, and 2001. Additional seeds are evaluated after primary execution succeeds. Missing outcomes receive the prespecified failure utility. Let $u_{sijr}$ be its operational utility, and let
\begin{equation}
\widehat q_{si}=\frac{1}{12}\sum_{j=1}^{12}\frac{1}{3}\sum_{r=1}^{3}u_{sijr}.
\end{equation}
The finite-reference promotion loss is
\begin{equation}
L_s^{(p)}=100a_{s0}\left(\max_i\widehat q_{si}-\sum_i\pi_{si}^{(p)}\widehat q_{si}\right),
\label{eq:followuploss}
\end{equation}
where $a_{s0}$ is baseline test accuracy. Missing seed outcomes receive the prespecified failure utility in ITT. Model-judged faithful sensitivities require all selection programs to execute on development and final data, every reference program to complete all three seeds, and the specified fidelity labels for every program. Each comparison uses at least two cards. Each sensitivity applies the policy distributions already saved for that card subset. Reference-size sensitivities retain this common support and use the first three, first six, or all twelve reference slots. Seed-zero references supply an additional sensitivity.

On the fixed ITT portfolio, two policies share the same empirical reference optimum. It cancels from their paired difference. Writing $d_{si}=\pi_{si}^{(p')}-\pi_{si}^{(p)}$ gives
\begin{equation}
L_s^{(p)}-L_s^{(p')}=100a_{s0}\sum_i d_{si}\widehat q_{si}.
\label{eq:policycontrast}
\end{equation}
The selection bank fixes weights $d_{si}$ before reference outcomes. If reference means are unbiased for utilities $\boldsymbol\mu_s$ under the declared sampling contract, with covariance $V_s$, linearity gives
\begin{align}
\E[L_s^{(p)}-L_s^{(p')}\mid\text{selection bank}]&=100a_{s0}\,\mathbf d_s^{\mathsf T}\boldsymbol\mu_s,\\
\Var(L_s^{(p)}-L_s^{(p')}\mid\text{selection bank})&=(100a_{s0})^2\mathbf d_s^{\mathsf T}V_s\mathbf d_s.
\end{align}
Thus reference noise contributes variance to the policy comparison, while the maximum's selection bias cancels. Absolute reference loss and winner agreement retain their dependence on the finite reference optimum. This motivates paired contrasts as primary policy endpoints and reference-size checks as supplementary diagnostics.

The retained-code comparison freezes an additional set of program weights. Within each world's selected card, the program with the highest observed selection score is retained; tied programs receive uniform weights. Its final three-seed utility is compared with $\widehat q_{si}$ for the same selected mechanism. Both actions use the same final training and evaluation data. The comparison measures the consequences of retaining a selected realization versus drawing a fresh realization of its mechanism. Development-based promotion loss also includes transfer from the development split to the final split. The same-data implementation-versus-seed diagnostic uses the first three fresh reference programs per card under the original matched design. All twelve fresh reference programs enter the variance analysis, using primary-seed outcomes and within-program seed variation.

The primary pooled analysis weights the eighteen tasks equally. It tests maximum-of-three minus mean-of-three loss and single-draw minus mean-of-five loss, with Holm adjustment across both contrasts and both execution processes in ITT. Exact paired sign-flip tests and 20,000 task-bootstrap resamples use seed 20261002. Separate summaries cover the twelve continuing tasks and six new tasks. Secondary analyses include all five policies, the original test-score bank on continuing tasks, the initial fixed-slot choices, and a zero-utility KEEP option. KEEP retains the frozen primary-seed baseline, consumes no extra implementation, and wins numerical ties. The plan, reuse provenance, failure records, and all analyses accompany the supplement.

\section{Execution and Review Protocol}
\label{app:execution}
\begin{table}[ht]
\centering
\caption{Task populations and held-out baseline accuracy. Identifiers fix the OpenML tasks and official splits. The initial prospective rule retains all twelve eligible tasks after exposure exclusions. The follow-up retains those tasks and adds all six eligible tasks under its expanded metadata envelope.}
\label{tab:tasks}
\small
\begin{tabular}{llrrrrrr}
\toprule
\rowcolor{ideablue!8} Cohort & Dataset & Task & Rows & Features & Classes & Test & Baseline (\%) \\
\midrule
Archived & balance-scale & 11 & 625 & 4 & 3 & 63 & 93.65 \\
Archived & mfeat-morphological & 18 & 2,000 & 6 & 10 & 200 & 63.50 \\
Archived & mfeat-zernike & 22 & 2,000 & 47 & 10 & 200 & 80.00 \\
Archived & spambase & 43 & 4,601 & 57 & 2 & 461 & 93.93 \\
Archived & satimage & 2074 & 6,430 & 36 & 6 & 643 & 92.85 \\
Archived & pc4 & 3902 & 1,458 & 37 & 2 & 146 & 91.78 \\
Archived & pc3 & 3903 & 1,563 & 37 & 2 & 157 & 87.26 \\
Archived & kc2 & 3913 & 522 & 21 & 2 & 53 & 81.13 \\
Archived & kc1 & 3917 & 2,109 & 21 & 2 & 211 & 87.20 \\
Archived & phoneme & 9952 & 5,404 & 5 & 2 & 541 & 89.46 \\
Archived & qsar-biodeg & 9957 & 1,055 & 41 & 2 & 106 & 87.74 \\
Archived & blood-transfusion & 10101 & 748 & 4 & 2 & 75 & 76.00 \\
Archived & segment & 146822 & 2,310 & 16 & 7 & 231 & 93.51 \\
Prospective & optdigits & 28 & 5,620 & 64 & 10 & 562 & 97.69 \\
Prospective & pendigits & 32 & 10,992 & 16 & 10 & 1,100 & 99.18 \\
Prospective & analcatdata\_authorship & 3549 & 841 & 70 & 4 & 85 & 98.82 \\
Prospective & pc1 & 3918 & 1,109 & 21 & 2 & 111 & 96.40 \\
Prospective & wall-robot-navigation & 9960 & 5,456 & 24 & 4 & 546 & 99.08 \\
Prospective & semeion & 9964 & 1,593 & 256 & 10 & 160 & 91.25 \\
Prospective & ozone-level-8hr & 9978 & 2,534 & 72 & 2 & 254 & 92.91 \\
Prospective & banknote-authentication & 10093 & 1,372 & 4 & 2 & 138 & 99.28 \\
Prospective & GesturePhase & 14969 & 9,873 & 32 & 5 & 988 & 68.42 \\
Prospective & steel-plates-fault & 146817 & 1,941 & 27 & 7 & 195 & 83.59 \\
Prospective & climate-simulation & 146819 & 540 & 18 & 2 & 54 & 96.30 \\
Prospective & wilt & 146820 & 4,839 & 5 & 2 & 484 & 97.52 \\
Follow-up & letter & 6 & 20,000 & 16 & 26 & 2,000 & 96.95 \\
Follow-up & isolet & 3481 & 7,797 & 617 & 26 & 780 & 95.77 \\
Follow-up & madelon & 9976 & 2,600 & 500 & 2 & 260 & 82.69 \\
Follow-up & cnae-9 & 9981 & 1,080 & 856 & 9 & 108 & 88.89 \\
Follow-up & har & 14970 & 10,299 & 561 & 6 & 1,030 & 99.71 \\
Follow-up & texture & 125922 & 5,500 & 40 & 11 & 550 & 99.09 \\
\bottomrule
\end{tabular}
\end{table}

\begin{figure}[ht]
\centering
\includegraphics[width=.95\textwidth]{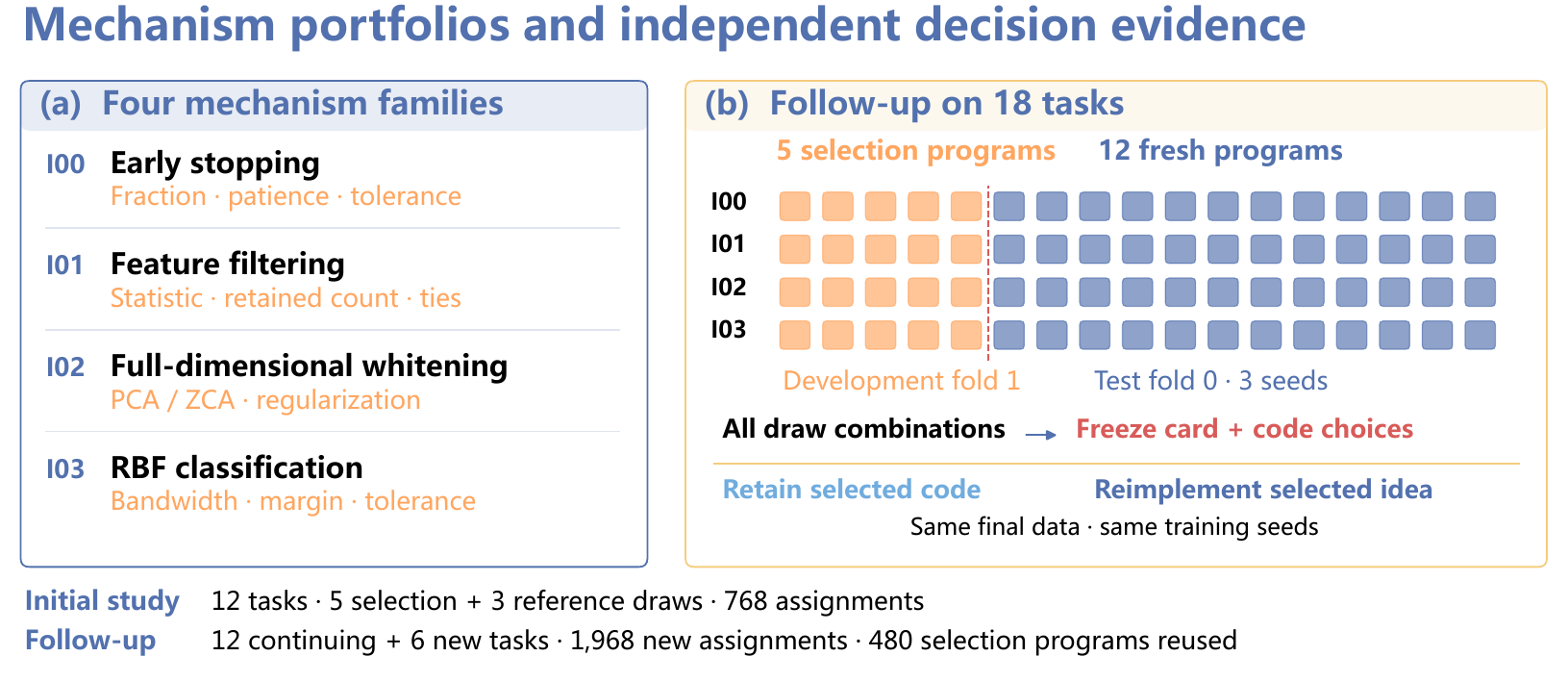}
\caption{Mechanism families and follow-up sampling. Each portfolio fixes four task-conditioned cards. Five saved programs per card supply development evidence; twelve fresh programs form the final reference. The twelve continuing tasks reuse their initial selection banks. Separate retained-code weights specify the artifact action.}
\label{fig:design}
\end{figure}

The archived cohort contains 13 tasks selected in a six-task initial cohort and a seven-task extension. The prospective cohort expands the metadata envelope to 12,000 rows and 256 numeric predictors and includes all eligible tasks remaining after the stated exposure exclusions. Exact task identifiers, visible baseline scores, split sizes, accepted cards, rejected portfolio attempts, and model prompts are included in the supplement.

The prospective portfolio uses task-conditioned specifications of four prespecified mechanism families. Pre-execution construction diagnostics identified duplicate or internally inconsistent free-form cards and motivated this controlled family design. Those diagnostic portfolios have no implementation outcomes and are retained separately. The final admission procedure requires both full-portfolio reviews to pass. Reviewers check each card's mechanism coherence, baseline distinction, compatibility, meaningful implementation freedom, and all pairwise distinctions. Implementation begins only after this admission record exists.

Evaluation runs on a local Linux CPU environment with 22 logical processors, an Intel Core Ultra 9 185H, and 15 GiB available physical memory. Each evaluator worker uses one thread. Local generation orchestration uses up to 24 task/process shards; held-out evaluation uses six concurrent shards. No GPU training is used.

The follow-up evaluates at most two task/process shards concurrently, each with six frozen-program workers. Each fit retains the 300-second allowance. Baseline preparation for isolet required a larger aggregate allowance for its nine-fold development summary, raised from 1,200 to 2,700 seconds before that task's cards were generated. Both campaigns use the same baseline predictor and per-fit allowance. Evaluation timeouts terminate the controller and its child processes; completed per-program records support resumption. Preparation amendments and the source used before evaluation are retained with the run.

Observed evaluation timeouts motivated a separate runtime diagnostic. Its plan was written after these failures and before the diagnostic evaluations. The inclusion rule selects every task with a timeout in its original primary or additional-seed evaluation. For each selected task, the diagnostic evaluates the complete five-selection-plus-twelve-reference bank under both processes, including previously successful programs, at six total concurrent fits. The diagnostic uses the saved code, data, seed zero, fidelity labels, selection probabilities, and 300-second allowance. Each new result is stored separately. The comparison uses the original seed-zero reference sensitivity as its matched target and also reports the original three-seed analysis on tasks without timeouts. The primary results and four adjusted tests use the original outcome matrix. This post-outcome diagnostic measures sensitivity to execution load.

All prospective model calls use DeepSeek Flash. Card generation and portfolio review use low reasoning effort and a 16,384-token output allowance. Both executors use the same frozen card and task packet. Bounded exposes capped model calls and permitted inspection/repair actions. Agentic uses a fresh Read/Edit SDK session and its requested turn limit. The session configuration is retained together with actual usage; an SDK option is recorded as a requested bound. Code-level protections preserve evaluator files and labels, and evaluation checks executable behavior under the declared compute policy.

Every prospective assignment receives two outcome-blind fidelity judgments through separate fixed Flash prompts. Their agreement measures consistency within that model. The primary label defines ITT fidelity penalties; a stricter conditional analysis requires both labels to be faithful. Constructed invalid and mismatched-card controls provide additional review diagnostics. These labels supply model-based evidence about specification fidelity. The saved code supports direct inspection of each concrete invariant. The archived cohort retains physical independence evidence for 311 of 312 assignments; ITT includes the assignment with one missing provider response identifier. The prospective cohort records each requested model role and every physical session.

The initial prospective judges agree on 374 of 384 Bounded assignments and 378 of 384 Agentic assignments. Both labels are faithful for 370 and 378 assignments. Each task/process shard supplies four mismatched-card and four empty-patch controls. All 96 mismatched-card controls receive the expected drift label, and all 96 empty-patch controls receive the expected invalid label. These constructed contrasts measure detection of clear violations; their accuracy supplies a separate diagnostic from agreement on ordinary implementations.

The 384 Agentic sessions use a median of 11 model turns, a mean of 12.0, and a maximum of 34. The recorded 20-turn SDK option therefore describes the request sent to the wrapper. Across the formal prospective collection, request ledgers contain 82 portfolio-generation/review calls, 1,672 Bounded implementation/review calls, 1,189 Agentic review calls, and 384 Agentic sessions. Every recorded requested model is \texttt{deepseek-flash}. The resource audit supplies uncached input, cached input, and output counts by process and role.

The follow-up judges agree on 968 of 984 Bounded assignments and 975 of 984 Agentic assignments. Both labels are faithful for 955 and 973 assignments. All 144 mismatched-card controls and 144 empty-patch controls receive their expected labels. Its generation records contain 45 portfolio-generation/review calls, 4,160 Bounded implementation/review calls, 2,951 Agentic review calls, and 984 Agentic SDK sessions. Every recorded requested model is \texttt{deepseek-flash}. The Agentic sessions use a median of 11 turns, a mean of 12.06, and a maximum of 27. The supplement preserves usage by role and the independent request/session records across both collections.

Execution supplies an additional validity check. On task 14969, Agentic card I00, follow-up reference slot 10 received two faithful labels and failed at runtime because its accuracy scorer had the wrong callable signature. The frozen code is retained, the failed attempt receives its ITT failure utility, and analyses of model-judged faithful executable programs exclude it. This example shows why semantic review and executable validity enter the audit separately.

The whole initial prospective cohort freezes code, review labels, and rerun assignments before evaluation. The evaluator computes selection-slot outcomes first and saves every policy's score vector and selected card. It then evaluates the reference slots and the prespecified saved-code seed reruns. The follow-up additionally freezes all development-based policy distributions before its final evaluation. Subsequent analyses read the recorded outcomes and apply the prespecified failure penalties.

\subsection{Reproduction materials}
The supplement supplies the analysis code, frozen protocol and configurations, task manifests, card and review records, assignment-level scores, saved-code rerun scores, policy selections, and figure sources. Offline scripts rebuild numerical summaries and figures from the retained evidence. Raw benchmark data can be obtained through the task identifiers from OpenML. The supplement identifies required packages and separates offline result reproduction from regenerating model calls. Machine-readable assignment tables link decisions to the programs they use.

The per-dataset metadata preserves OpenML's creator, citation, and license fields with their primary-source endpoints. All 31 returned license fields read ``Public''; these provider labels are retained verbatim. NumPy, SciPy, and scikit-learn retain their BSD distribution notices, and Matplotlib retains its own license notice. DeepSeek, the Agentic SDK, and Microsoft YaHei are externally accessed assets. The supplement includes derived vector figures and excludes benchmark arrays, provider credentials, and proprietary font binaries.

\section{Semantic and Decision Sensitivities}
\label{app:robustness}
The archived conservative screen retains 38 cards. Its model-judged faithful execution ICC is 0.617 [0.230, 0.857] for Bounded and 0.443 [0.085, 0.826] for Agentic. Complete-card decision support covers 11 and eight tasks. Mean accuracy loss is 0.287 and 0.721 points for independent implementations, compared with 0.105 and 0.118 points for saved-code reruns. Their paired differences are 0.182 [0.050, 0.336] and 0.602 [0.044, 1.525] points. Both semantic screens produce positive excesses that combine selection behavior with unequal reference floors.

The archived morphological and Zernike tasks represent feature views of shared handwritten digits \citep{duin1998multiple}. A supplementary bootstrap resamples these views together and preserves the original point estimates and task weights. The construct-admitted, model-judged faithful Bounded variance analysis then has 12 source clusters and ICC 0.772 [0.531, 0.891]. Agentic variance filtering retains one view. Both matched loss comparisons also retain at most one view after complete-card filtering, so their cluster counts are unchanged. Source-cluster intervals and counts accompany the offline analysis outputs.

The original aligned-index ITT audit gives winner-disagreement rates of 25.6\% and 43.6\%. Complete enumeration gives 22.5\% and 45.3\%, while disjoint winner sets give 13.7\% and 32.5\%. Accuracy-equivalent ITT loss averages 0.271 and 0.661 points. These quantities include operational penalties. The model-judged faithful population losses in the main text use absolute accuracy points.

\begin{figure*}[!t]
\centering
\includegraphics[width=\textwidth]{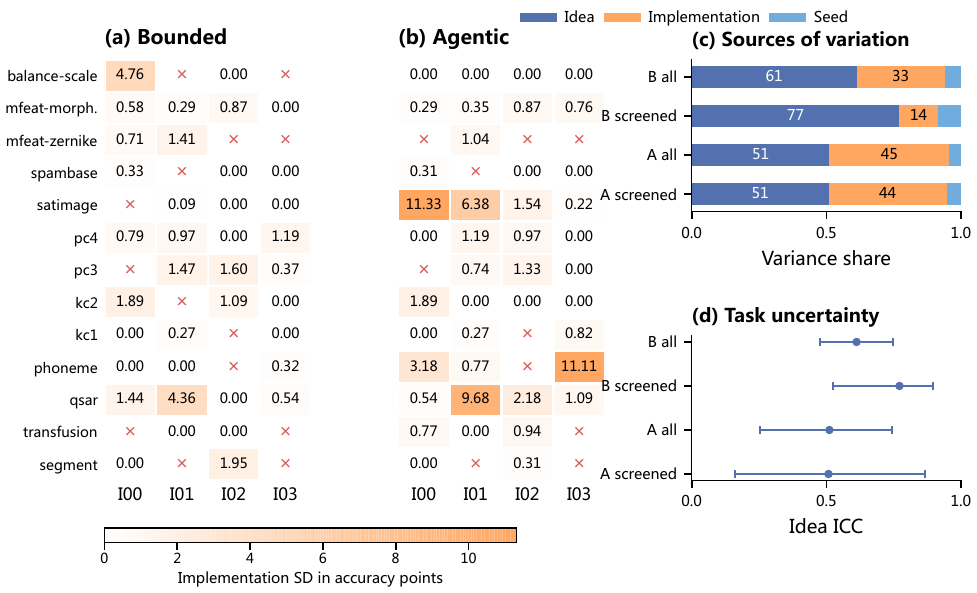}
\caption{Archived variation at card and population levels. Panels (a--b) show standard deviations across model-judged faithful executable implementations of each semantically admitted card, in accuracy points. A cross denotes a filtered card or fewer than two eligible implementations. Card labels are local to each task. Panels (c--d) compare all assigned attempts with the admitted, model-judged faithful population; B and A denote Bounded and Agentic; screened denotes construct-admitted, model-judged faithful programs. Error bars are 95\% task-bootstrap intervals. The heatmaps retain all eligible cells and their zero values.}
\label{fig:variation}
\end{figure*}

\begin{figure}[ht]
\centering
\includegraphics[width=.95\textwidth]{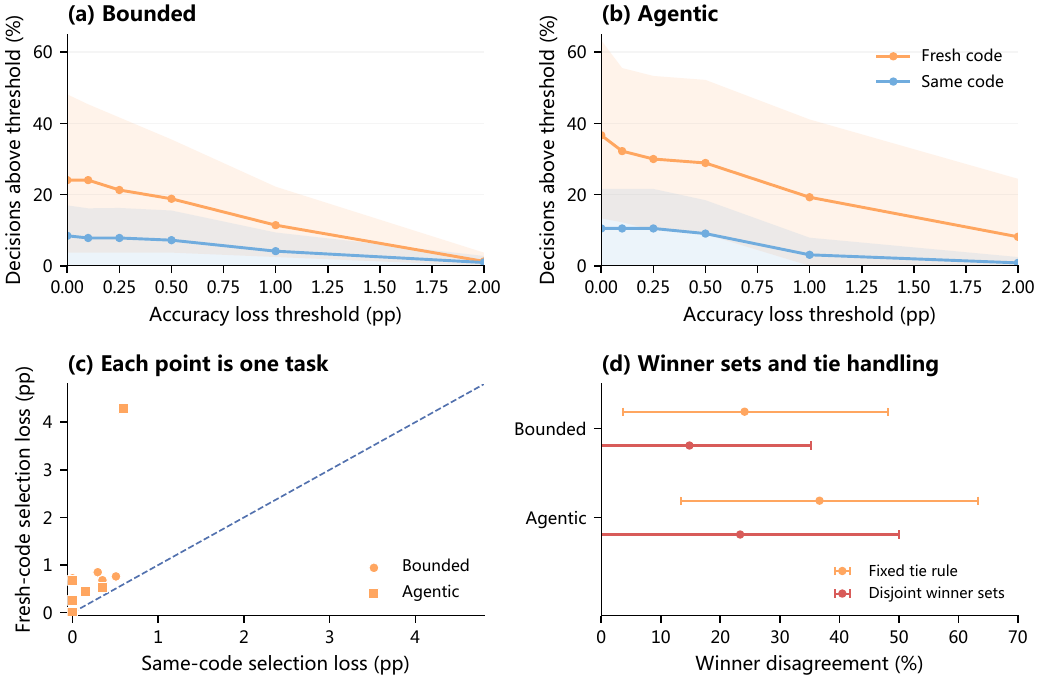}
\caption{Exploratory archived decision diagnostics on admitted, model-judged faithful complete cards. Independent-code and saved-code references have different noise floors and conditional targets. Panels retain their loss profiles, winner-set overlap, and task-level comparisons. Excess reference loss therefore combines these differences with selection behavior. Error bands and intervals resample tasks.}
\label{fig:decisions}
\end{figure}

\begin{table}[ht]
\centering
\caption{Prospective population coverage and idea ICC. Variance uses all retained implementations; policy evaluation requires all eight eligible draws per retained card. Every variance estimate covers 48 cards on 12 tasks.}
\label{tab:prospective_coverage}
\small
\begin{tabular}{llrrrr}
\toprule
\rowcolor{ideablue!8} Process & Population & Rows & Policy cards & Policy tasks & ICC [95\% CI] \\
\midrule
Bounded & ITT & 384 & 48 & 12 & 0.564 [0.497, 0.840] \\
Bounded & Model-judged & 380 & 44 & 12 & 0.591 [0.525, 0.932] \\
Bounded & Both-review & 370 & 37 & 11 & 0.590 [0.523, 0.932] \\
Agentic & ITT & 384 & 48 & 12 & 0.851 [0.719, 0.942] \\
Agentic & Model-judged & 384 & 48 & 12 & 0.851 [0.719, 0.942] \\
Agentic & Both-review & 378 & 43 & 12 & 0.851 [0.718, 0.943] \\
\bottomrule
\end{tabular}
\end{table}

\subsection{Initial prospective results}
The initial prospective cohort retains 380 model-judged faithful executable Bounded assignments and all 384 Agentic assignments. ITT includes the four invalid Bounded attempts. All 764 eligible programs complete their three seed evaluations, and every implementation has verified physical independence. The model-judged faithful population ICC is 0.591 [0.525, 0.932] for Bounded and 0.851 [0.719, 0.942] for Agentic. Absolute implementation and seed standard deviations, together with score-invariant rerun counts, appear in Table~\ref{tab:absolute}. Requiring both judges to label an implementation faithful gives ICCs of 0.590 and 0.851, with coverage in Appendix~\ref{app:robustness}.

\begin{table*}[!t]
\centering
\caption{Initial prospective selection loss with 95\% task-bootstrap intervals. ITT reports accuracy-equivalent utility points for all 48 cards. Model-judged faithful complete-card results report accuracy points for 44 Bounded and 48 Agentic cards. Each analysis covers 12 tasks; conditional policies use retained cards.}
\label{tab:policies}
\small
\begin{tabular}{lrrrr}
\toprule
\rowcolor{ideablue!8} & \multicolumn{2}{c}{ITT utility loss} & \multicolumn{2}{c}{Model-judged accuracy loss} \\
Policy & Bounded & Agentic & Bounded & Agentic \\
\midrule
Single draw & 0.258 [0.000, 0.752] & 0.000 [0.000, 0.000] & 0.023 [0.000, 0.069] & 0.000 [0.000, 0.000] \\
Mean of 3 & 0.291 [0.000, 0.785] & 0.000 [0.000, 0.000] & 0.023 [0.000, 0.069] & 0.000 [0.000, 0.000] \\
Maximum of 3 & 0.645 [0.000, 1.425] & 0.206 [0.000, 0.617] & 0.101 [0.000, 0.304] & 0.206 [0.000, 0.617] \\
Mean of 5 & 0.291 [0.000, 0.785] & 0.000 [0.000, 0.000] & 0.023 [0.000, 0.069] & 0.000 [0.000, 0.000] \\
Maximum of 5 & 0.678 [0.066, 1.437] & 0.206 [0.000, 0.617] & 0.101 [0.000, 0.304] & 0.206 [0.000, 0.617] \\
\bottomrule
\end{tabular}
\end{table*}

\begin{figure*}[!t]
\centering
\includegraphics[width=\textwidth]{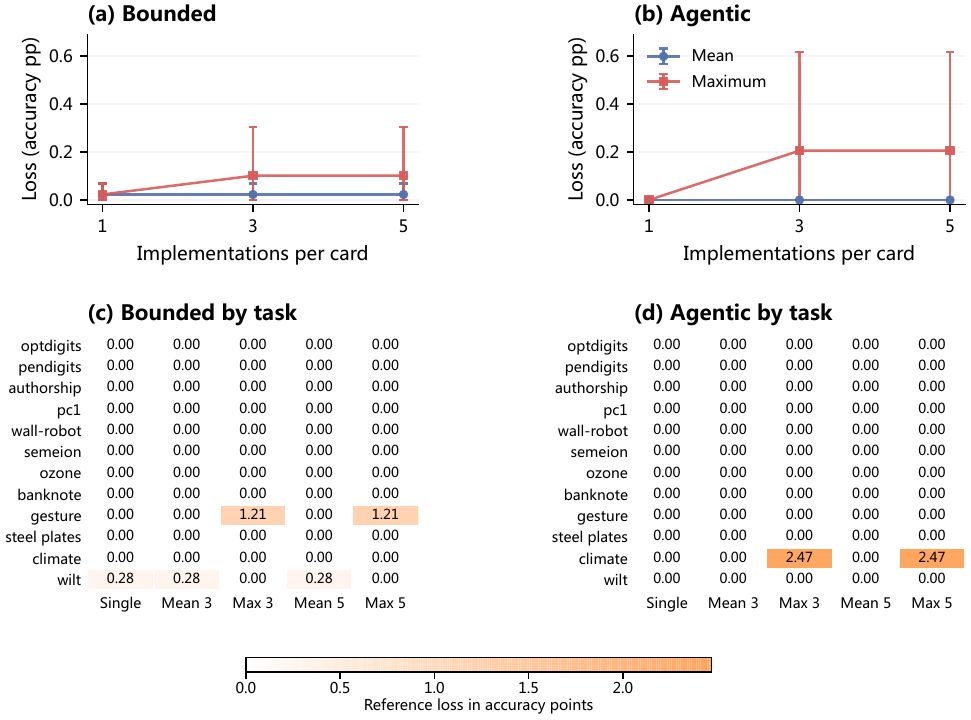}
\caption{Initial prospective selection on the model-judged faithful complete-card population. Panels (a--b) show loss against the three fresh reference implementations, with 95\% task-bootstrap intervals. Single-draw points are shared by both curves; coincident means are retained. Panels (c--d) display every task's loss under all five policies. The maximum-based errors concentrate on different tasks in the two processes. The zero-loss Agentic single-draw and mean policies describe this particular frozen selection/reference partition.}
\label{fig:prospective}
\end{figure*}

Substantial implementation variation coexists with stable selection in this portfolio. On complete model-judged faithful cards, the prespecified single draw loses 0.023 accuracy points under Bounded execution and zero under Agentic execution. Means of three or five draws attain the same losses. The maximum policies lose 0.101 and 0.206 points. \tabref{tab:policies} also reports the full ITT results. Its Bounded single-draw loss is 0.258 utility points, compared with 0.291 for either mean policy and 0.645/0.678 for the three/five-draw maximum. Policy differences concentrate on a few tasks, and all four prespecified ITT contrasts have exact paired sign-flip $p\geq0.5$ in both processes. The observed maximum-policy losses describe the matrix with task-level confidence bounds.

The matched three-draw control compares empirical-reference losses under different reference variances and program-conditional targets. Fresh-code and saved-code losses are 0.085 and 0.089 accuracy points for Bounded, and 0.320 and 0.088 for Agentic. Their paired excess losses are $-0.004$ [$-0.173$, 0.176] and 0.231 [$-0.015$, 0.515]. Shared-reference policy endpoints accompany these two-reference diagnostics. In the prospective model-judged faithful population, reference top-two margins average 1.43 and 1.39 accuracy points, and most fixed-policy selections agree with that reference.

The prospective both-review faithful sensitivity requires both judges to label every retained implementation faithful. Complete eight-draw policy support then covers 37 cards on 11 Bounded tasks and 43 cards on 12 Agentic tasks. Bounded single, mean-of-three, mean-of-five, and maximum-policy losses are 0.030, 0.025, 0.025, and 0.307 accuracy points. Agentic losses are less than 0.001, 0.113, 0.015, and 0.098 points. Both maximum budgets yield the same value within each process. The four paired contrasts have $p\geq0.25$ for Bounded and $p\geq0.5$ for Agentic. These changes illustrate how the conditional card population affects policy comparisons. The release supplies every population's estimates, intervals, paired task differences, and exact sign-flip values.

\begin{figure}[ht]
\centering
\includegraphics[width=\textwidth]{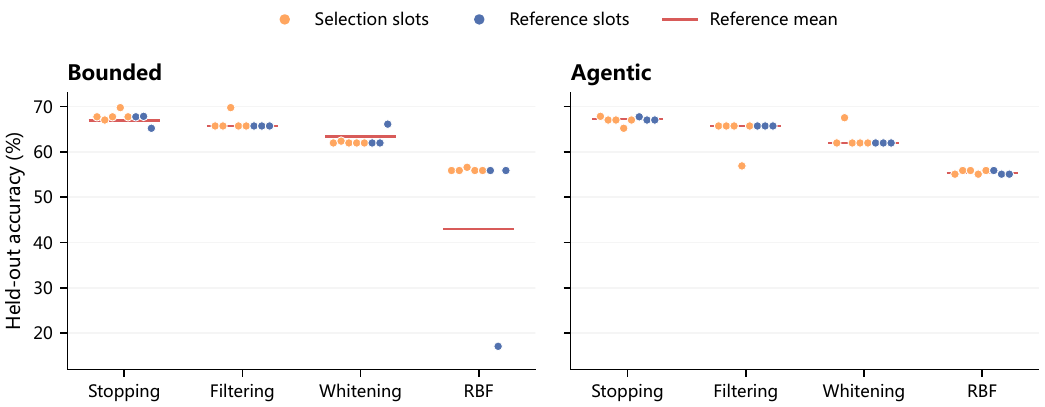}
\caption{All eight initial prospective implementations per card on the gesture task. Horizontal offsets separate the frozen draw slots; colors identify selection and reference roles. Every plotted program is model-judged faithful and executable. The red segments mark the means of the three reference slots. Large variability in a lower-ranked card can coexist with a stable leading mechanism. This case illustrates the distinction between a portfolio's variance and its promotion decisions.}
\label{fig:realizations}
\end{figure}

The prospective Bounded variance is influenced by a high-variation RBF card on the gesture task, whose eight model-judged faithful implementations have an accuracy standard deviation of 13.74 points. Its reference mean is below the leading early-stopping mechanism. In the Agentic process, the largest card-level standard deviation is 3.11 points for feature filtering on that task. Figure~\ref{fig:realizations} retains every realization of this portfolio. Variance describes the full portfolio; promotion loss measures its selected candidate.

\subsection{Follow-up evidence}
\label{app:followup-results}
The follow-up contributes 1,968 assigned attempts. Primary evaluation succeeds for 948 of 984 Bounded attempts and 951 of 984 Agentic attempts. Nine and one attempts are invalid before hidden evaluation. Runtime failures comprise 25 and 30 primary timeouts, respectively, plus two program errors in each process. All 55 primary timeouts occur on isolet. Its additional-seed evaluations produce seven and four timeouts. There are 944 and 948 complete three-seed records. Complete and partial records together contain 2,837 and 2,849 observed seed scores. Missing outcomes retain their prespecified ITT utility.

The reference analysis uses 864 assigned programs per process. Table~\ref{tab:followup-coverage} separates program-level variance coverage from the stricter complete-bank support needed for policy selection. The latter requires all five selection programs and all twelve reference programs for each retained card to satisfy execution and fidelity requirements. Both-review support additionally requires both judges to return the faithful label. Filtering defines the comparison target.

\begin{table}[ht]
\centering
\caption{Follow-up coverage and variance. All variance estimates span eighteen tasks. The final column gives the tasks and cards supporting complete-bank policy comparisons. Model-judged and Both-review denote the primary and dual model-fidelity requirements. Table~\ref{tab:absolute} supplies absolute components and task-bootstrap intervals.}
\label{tab:followup-coverage}
\small
\begin{tabular}{llrrlr}
\toprule
\rowcolor{ideablue!8} Process & Population & Programs & Cards & ICC [95\% interval] & Policy tasks/cards \\
\midrule
Bounded & ITT & 864 & 72 & 0.826 [0.658, 0.881] & 18 / 72 \\
Bounded & Model-judged & 833 & 71 & 0.880 [0.772, 0.933] & 16 / 52 \\
Bounded & Both-review & 819 & 71 & 0.919 [0.772, 0.975] & 15 / 44 \\
Agentic & ITT & 864 & 72 & 0.907 [0.739, 0.961] & 18 / 72 \\
Agentic & Model-judged & 835 & 70 & 0.921 [0.728, 0.970] & 18 / 65 \\
Agentic & Both-review & 827 & 70 & 0.921 [0.730, 0.970] & 18 / 59 \\
\bottomrule
\end{tabular}
\end{table}

\begin{table}[ht]
\centering
\caption{All five follow-up promotion policies. The first three rows within each process use ITT and report accuracy-equivalent utility points. The two conditional rows use accuracy points on their retained task/card populations. Parentheses give task counts. Figure~\ref{fig:extension-policies} supplies the primary uncertainty intervals; complete intervals and paired task differences for every population accompany the analysis files.}
\label{tab:followup-policies}
\small
\begin{tabular}{llrrrrr}
\toprule
\rowcolor{ideablue!8} Process & Population & Single & Mean 3 & Max 3 & Mean 5 & Max 5 \\
\midrule
Bounded & Continuing 12 & 0.345 & 0.229 & 0.412 & 0.229 & 0.406 \\
Bounded & New 6 & 0.765 & 0.958 & 0.166 & 0.997 & 0.095 \\
Bounded & Pooled 18 & 0.485 & 0.472 & 0.330 & 0.485 & 0.303 \\
Bounded & Model-judged (16) & 0.378 & 0.303 & 0.255 & 0.402 & 0.246 \\
Bounded & Both-review (15) & 0.156 & 0.056 & 0.231 & 0.016 & 0.262 \\
Agentic & Continuing 12 & 0.434 & 0.406 & 0.432 & 0.397 & 0.443 \\
Agentic & New 6 & 0.356 & 0.147 & 0.445 & 0.023 & 0.543 \\
Agentic & Pooled 18 & 0.408 & 0.320 & 0.436 & 0.273 & 0.476 \\
Agentic & Model-judged (18) & 0.274 & 0.193 & 0.365 & 0.061 & 0.499 \\
Agentic & Both-review (18) & 0.273 & 0.193 & 0.365 & 0.061 & 0.499 \\
\bottomrule
\end{tabular}
\end{table}

For Agentic, the model-judged faithful single-minus-mean-of-five contrast is 0.214 [0.026, 0.440] accuracy points with unadjusted exact sign-flip $p=0.03125$. The corresponding Bounded contrast is $-0.024$ [$-0.380$, 0.283], $p=0.828$. Both-review filtering gives 0.140 [0.012, 0.357], $p=0.094$, for Bounded and 0.212 [0.024, 0.438], $p=0.03125$, for Agentic. These are secondary conditional comparisons. The primary four-test Holm family uses the complete ITT portfolios and has adjusted values 1.0, 1.0, 1.0, and 0.5625.

The Bounded CNAE-9 portfolio illustrates how filtering changes the comparison. Mean-of-five loses 5.324 points under both ITT and model-judged faithful evaluation. One RBF reference program receives faithful and partial labels from the two judges, so the both-review faithful complete-bank rule excludes that card. The remaining early-stopping and whitening cards both have negative reference utility. Early stopping then wins their comparison, giving zero promotion loss. The reduction reflects a changed candidate set and reference optimum. Baseline-relative harm supplies complementary information about the selected candidate's absolute value.

\begin{table}[ht]
\centering
\caption{Reference-size and seed sensitivities for pooled ITT promotion loss. Every row retains the same eighteen tasks, complete four-card portfolios, and frozen selection distributions. Reference subsets use the first assigned slots. Values are accuracy-equivalent utility points against the stated finite reference.}
\label{tab:followup-reference}
\small
\begin{tabular}{llrrrrr}
\toprule
\rowcolor{ideablue!8} Process & Reference per card & Single & Mean 3 & Max 3 & Mean 5 & Max 5 \\
\midrule
Bounded & 3 programs / 3 seeds & 0.548 & 0.393 & 0.479 & 0.326 & 0.443 \\
Bounded & 6 programs / 3 seeds & 0.470 & 0.430 & 0.318 & 0.415 & 0.284 \\
Bounded & 12 programs / 3 seeds & 0.485 & 0.472 & 0.330 & 0.485 & 0.303 \\
Bounded & 12 programs / seed 0 & 0.668 & 0.628 & 0.525 & 0.637 & 0.499 \\
Agentic & 3 programs / 3 seeds & 0.421 & 0.312 & 0.427 & 0.261 & 0.473 \\
Agentic & 6 programs / 3 seeds & 0.443 & 0.333 & 0.493 & 0.288 & 0.567 \\
Agentic & 12 programs / 3 seeds & 0.408 & 0.320 & 0.436 & 0.273 & 0.476 \\
Agentic & 12 programs / seed 0 & 0.656 & 0.553 & 0.742 & 0.503 & 0.796 \\
\bottomrule
\end{tabular}
\end{table}

\begin{table}[ht]
\centering
\caption{Selection-signal sensitivities on the twelve continuing tasks. All rows use the new twelve-program, three-seed reference. The original test-score bank was observed in the initial study. Fixed-slot rows preserve its original selected labels exactly. All-combination rows average every permitted subset of the same five-program bank. Development selection uses a separate fold and therefore also includes split transfer. Values are ITT accuracy-equivalent utility points.}
\label{tab:followup-signal}
\small
\begin{tabular}{llrrrrr}
\toprule
\rowcolor{ideablue!8} Process & Selection evidence & Single & Mean 3 & Max 3 & Mean 5 & Max 5 \\
\midrule
Bounded & Original test / fixed slots & 0.079 & 0.176 & 0.232 & 0.176 & 0.329 \\
Bounded & Original test / all combinations & 0.205 & 0.156 & 0.223 & 0.176 & 0.329 \\
Bounded & Development / all combinations & 0.345 & 0.229 & 0.412 & 0.229 & 0.406 \\
Agentic & Original test / fixed slots & 0.069 & 0.069 & 0.116 & 0.069 & 0.116 \\
Agentic & Original test / all combinations & 0.217 & 0.069 & 0.207 & 0.069 & 0.116 \\
Agentic & Development / all combinations & 0.434 & 0.406 & 0.432 & 0.397 & 0.443 \\
\bottomrule
\end{tabular}
\end{table}

\begin{table}[ht]
\centering
\caption{Matched same-data decision controls in the follow-up. The first three fresh reference programs supply the implementation comparison. The saved-code comparison fixes a program per card and varies its three seeds. Both use the same eligible cards and two reference draws per card. The paired difference includes the distinct reference variances and conditional targets. Values are accuracy points with 95\% task-bootstrap intervals. The model-judged faithful control is also the valid-code control reported for ITT.}
\label{tab:followup-controls}
\small
\begin{tabular}{llrlll}
\toprule
\rowcolor{ideablue!8} Process & Fidelity & Tasks & Implementation & Saved code & Paired difference \\
\midrule
Bounded & Model-judged & 18 & 0.213 [0.051, 0.419] & 0.070 [0.008, 0.155] & 0.143 [-0.046, 0.357] \\
Bounded & Both-review & 18 & 0.210 [0.049, 0.416] & 0.072 [0.010, 0.157] & 0.138 [-0.049, 0.350] \\
Agentic & Model-judged & 18 & 0.129 [0.022, 0.290] & 0.075 [0.014, 0.145] & 0.054 [-0.064, 0.187] \\
Agentic & Both-review & 18 & 0.061 [0.000, 0.134] & 0.087 [0.014, 0.176] & -0.026 [-0.140, 0.081] \\
\bottomrule
\end{tabular}
\end{table}

\clearpage
\subsection{Runtime sensitivity}

\label{app:runtime-results}

The error-type rule selects isolet alone. Its complete banks contain 136 assigned programs, including one invalid Bounded assignment. At six total concurrent fits, 54 Bounded and 55 Agentic programs complete, compared with 42 and 38 under the original seed-zero evaluation. The remaining evaluation failures number 13 and 13. All 80 programs successful under both conditions have identical integer correct counts. This comparison identifies runtime sensitivity in the operational outcomes while preserving the original primary matrix.

\begin{table}[ht]

\centering

\caption{Runtime sensitivities for all five promotion policies. The first two rows compare seed-zero references on eighteen tasks; only isolet changes evaluation conditions. The final row retains original three-seed outcomes on the seventeen tasks without timeouts. Values are ITT accuracy-equivalent utility points.}

\label{tab:runtime-policies}

\small

\begin{tabular}{llrrrrr}

\toprule

\rowcolor{ideablue!8} Process & Evaluation & Single & Mean 3 & Max 3 & Mean 5 & Max 5 \\

\midrule

Bounded & Original / seed 0 & 0.668 & 0.628 & 0.525 & 0.637 & 0.499 \\

Bounded & Reduced concurrency / seed 0 & 0.651 & 0.611 & 0.525 & 0.637 & 0.499 \\

Bounded & No-timeout tasks / 3 seeds & 0.464 & 0.398 & 0.349 & 0.514 & 0.320 \\

Agentic & Original / seed 0 & 0.656 & 0.553 & 0.742 & 0.503 & 0.796 \\

Agentic & Reduced concurrency / seed 0 & 0.656 & 0.553 & 0.742 & 0.503 & 0.796 \\

Agentic & No-timeout tasks / 3 seeds & 0.432 & 0.339 & 0.462 & 0.289 & 0.504 \\

\bottomrule

\end{tabular}

\end{table}

A second runtime analysis was specified after observing the primary results and 66 diagnostic outcomes. It evaluates retained code and fresh implementation at seed zero for both actions, using the same frozen choice distributions. Both isolet banks receive the planned reduced-concurrency outcomes; other tasks retain their original seed-zero outcomes. The addendum requires no additional training. Table~\ref{tab:runtime-retention} includes every policy; its intervals resample tasks without additional hypothesis tests.

\begin{table}[ht]

\centering

\caption{Retained-code utility minus fresh-implementation utility under matched seed-zero evaluation. Both rows per process contain eighteen tasks. Values are accuracy-equivalent utility points. All five policies retain their original development-based selection distributions.}

\label{tab:runtime-retention}

\small

\begin{tabular}{llrrrrr}

\toprule

\rowcolor{ideablue!8} Process & Evaluation & Single & Mean 3 & Max 3 & Mean 5 & Max 5 \\

\midrule

Bounded & Original & 0.263 & 0.246 & 0.390 & 0.305 & 0.336 \\

Bounded & Reduced concurrency & 0.270 & 0.256 & 0.390 & 0.305 & 0.336 \\

Agentic & Original & 0.212 & 0.333 & 0.395 & 0.432 & 0.508 \\

Agentic & Reduced concurrency & 0.212 & 0.333 & 0.395 & 0.432 & 0.508 \\

\bottomrule

\end{tabular}

\end{table}

Mean-of-five selected-code differences are 0.305 [$-0.085$, 0.767] for Bounded and 0.432 [0.039, 1.105] for Agentic under both runtime conditions. These comparisons measure sensitivity to evaluation load using the same finite reference. The primary tests use the original evaluation records.

\begin{figure}[ht]
\centering
\includegraphics[width=.95\textwidth]{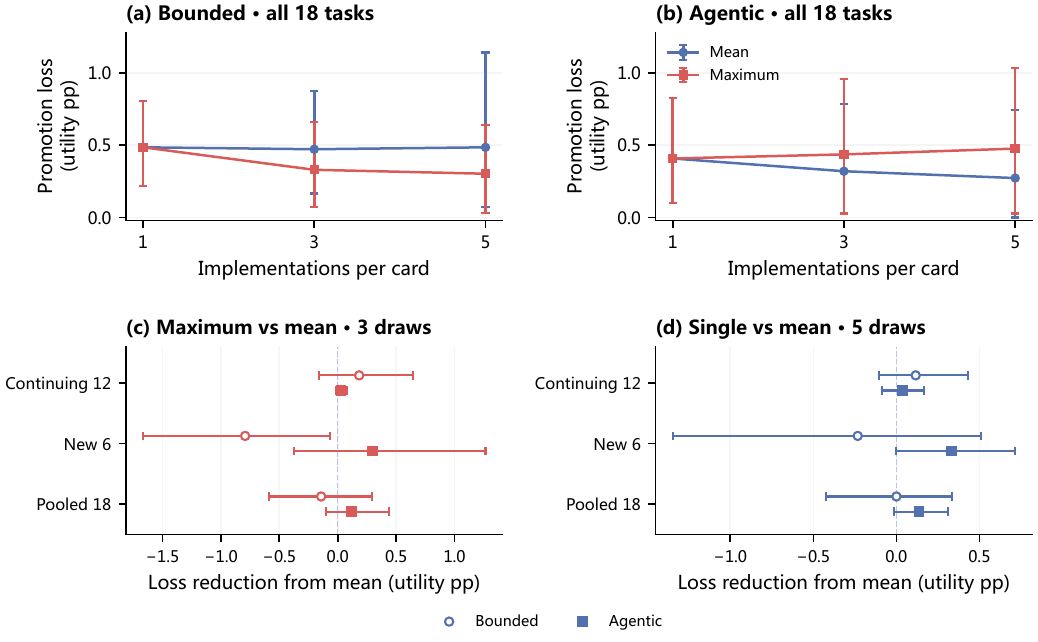}
\caption{Complete follow-up policy losses and primary contrasts. Positive contrast values favor the mean policy. Intervals resample complete tasks within the stated groups. The primary pooled ITT Holm values are 1, 1, 1, and 0.5625. Absolute losses include finite-reference noise and development-to-final transfer.}
\label{fig:extension-policies}
\end{figure}

\begin{figure}[ht]
\centering
\includegraphics[width=.9\textwidth]{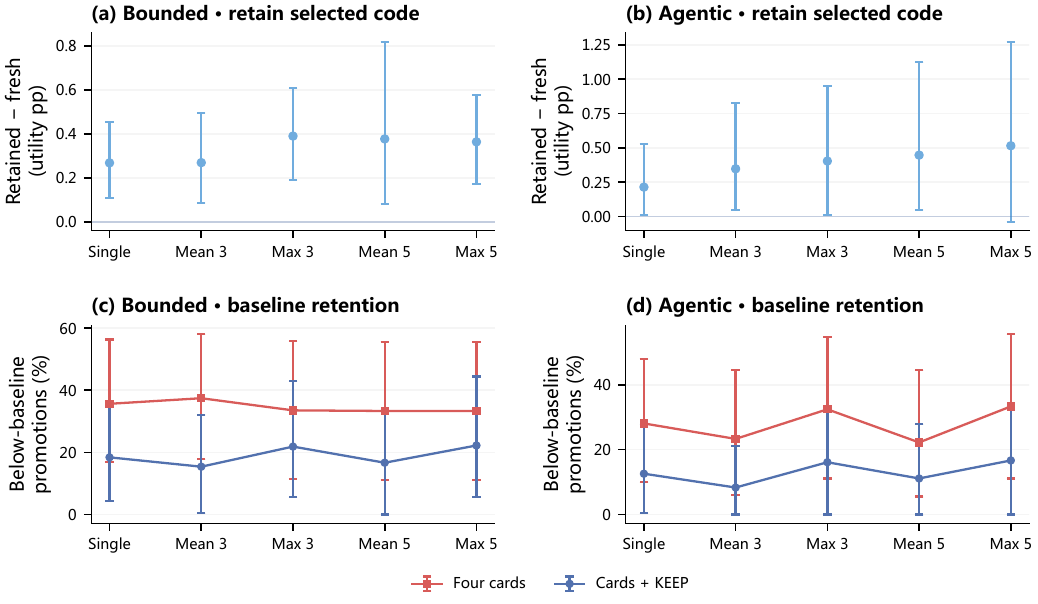}
\caption{Retained-code and baseline-KEEP sensitivities for all five policies. Retained-code intervals resample tasks in ITT. KEEP retains the frozen baseline at zero normalized utility and wins selection ties. Allowing KEEP can reduce negative-reference promotions while declining positive-reference candidates. These intervals describe secondary comparisons of the two research actions.}
\label{fig:extension-actions}
\end{figure}

\clearpage
\section{Analysis Timing and Additional Diagnostics}
\label{app:diagnostics}
\subsection{Design timing and analysis populations}
\label{app:timeline}
The prospective plans are local prespecifications documented by protocol files, manifests, and freeze records. The follow-up responds to the initial results; its primary four-test ITT family was fixed before new final outcomes. The six new tasks were collected prospectively, while their separate post-outcome analyses were specified after results. The analysis inventory records this sequence.

\begin{table}[ht]
\centering\small
\caption{Timing of design and analysis decisions. Offline diagnostics use saved outcomes and frozen programs. Separately recorded runtime and SVC checks add CPU evaluations; no diagnostic adds model calls.}
\label{tab:timing}
\begin{tabular}{p{.26\textwidth}p{.28\textwidth}p{.37\textwidth}}
\toprule
\rowcolor{ideablue!8} Component & Evidence available at design & Status and interpretation \\
\midrule
Archived semantic screens and audit & Implementations already generated; judges blinded to scores & Post-implementation admission, exploratory comparisons \\
Initial prospective cards and policies & Task metadata and visible baseline; no implementation outcomes & Admission before code; code/reviews/seeds frozen cohort-wide before outcomes \\
Follow-up portfolio and primary tests & Initial twelve-task results observed & Locally prespecified follow-up; reuse of 480 selection programs documented \\
Selected code versus fresh implementation & Initial results observed; new final outcomes sealed & Prespecified secondary endpoint; development-based weights frozen before new final evaluation \\
Six added tasks & Metadata and baseline; no earlier implementation outcomes & Prospective collection; secondary subgroup estimates \\
Runtime-load diagnostic & Primary timeouts observed & Post-outcome plan, separate evaluations, original outcomes preserved \\
Cross-split transport, item uncertainty, Gaussian calibration & All original outcomes observed & Post-outcome offline diagnostics with an explicit analysis plan \\
Premium robustness, concentration, headroom and bank decomposition & All original outcomes observed & Post-outcome diagnostics; original endpoint and four-test family preserved \\
\bottomrule
\end{tabular}
\end{table}

The initial cohort uses all remaining eligible tasks with complete numeric predictors, at most 12,000 rows, 256 predictors, ten classes, and imbalance at most 20. Exposure exclusions remove archived tasks and related multiple-features handwriting views. The follow-up expands the envelope to 20,000 rows, 900 predictors, and 26 classes, adding all six eligible new tasks. These rules permit high-baseline and low-headroom tasks. Task manifests preserve the full screen and exact identifiers.

\subsection{Absolute variation and seed-score invariance}
The nested moment model estimates between-card, implementation, and seed variance after task centering. We denote the between-card effect by $A_{si}$, so $Y_{sijk}=\mu_s+A_{si}+M_{sij}+R_{sijk}$ and $\rho_{\rm card}=\sigma_A^2/(\sigma_A^2+\sigma_M^2+\sigma_R^2)$. I00--I03 identify the cards. ICC values use normalized gains. The absolute estimates below transform those outcomes to accuracy points before estimating moments. The initial and follow-up populations require primary model-judged fidelity and complete seeds; the archived population additionally requires construct admission and at least two retained cards per task. The machine-readable records include raw negative component estimates; reported standard deviations use components bounded below by zero.

\begin{table}[ht]
\centering\small
\caption{Absolute standard deviations in accuracy points and exactly equal three-seed scores. Equality concerns aggregate accuracy at the three observed seeds. Deterministic predictions or code require additional evidence. Intervals resample complete tasks.}
\label{tab:absolute}
\begin{tabular}{llllrr}
\toprule
\rowcolor{ideablue!8} Cohort & Process & Implementation SD & Seed SD & Equal-score programs & Fraction \\
\midrule
Archived & Bounded & 1.095 [0.000, 1.720] & 0.752 [0.457, 1.000] & 77/111 & 69.4\% \\
Archived & Agentic & 2.649 [0.000, 3.855] & 0.820 [0.469, 1.091] & 81/111 & 73.0\% \\
Initial & Bounded & 2.160 [0.382, 3.598] & 0.429 [0.229, 0.588] & 316/380 & 83.2\% \\
Initial & Agentic & 0.926 [0.309, 1.368] & 0.427 [0.229, 0.580] & 314/384 & 81.8\% \\
Follow-up & Bounded & 1.169 [0.461, 1.701] & 0.459 [0.290, 0.599] & 647/833 & 77.7\% \\
Follow-up & Agentic & 0.983 [0.579, 1.401] & 0.451 [0.292, 0.585] & 655/835 & 78.4\% \\
\bottomrule
\end{tabular}
\end{table}

The evaluated scikit-learn version is 1.9.0. Histogram gradient boosting with automatic early stopping uses the training-size rule to determine whether a validation split is active. The early-stopping card explicitly enables validation monitoring, so smaller tasks can still use randomized splitting. Other permitted choices also affect seed sensitivity. Reporting absolute standard deviations and equal-score counts makes these controls interpretable even when some programs have zero observed seed variance. The archived illustrative gains correspond to five, one, and four test items in absolute magnitude. Small test sets make such score steps visible; the paired item analysis models their sampling contribution.

\subsection{Task concentration and test resolution}
Positive task differences favor the mean policy in each named contrast. The columns P/T/N count positive, tied, and negative task differences. Minimum $p$ is the two-sided sign-flip lower bound determined by nonzero support. Primary pooled follow-up intervals retain the original bootstrap seed and outcomes. Subgroup intervals also retain the original follow-up computation; post-outcome diagnostics add task counts and leave-one-task-out ranges. The new-task four-test Holm values are 0.50, 1, 1, and 1 for Bounded Max3--Mean3, Bounded Single--Mean5, Agentic Max3--Mean3, and Agentic Single--Mean5.

\begin{table}[ht]
\centering\small
\caption{Follow-up task support for the two primary contrast definitions. All-18 tests form the primary Holm family; new-six tests are secondary.}
\label{tab:contrast-support}
\begin{tabular}{llllrrr}
\toprule
\rowcolor{ideablue!8} Group & Process & Contrast & Difference [95\% CI] & P/T/N & Raw $p$ & Min. $p$ \\
\midrule
All 18 & Bounded & Max3--Mean3 & -0.142 [-0.591, 0.292] & 3/9/6 & 0.5898 & 0.0039 \\
All 18 & Bounded & Single--Mean5 & 0.000 [-0.425, 0.337] & 5/8/5 & 1.0000 & 0.0020 \\
All 18 & Agentic & Max3--Mean3 & 0.116 [-0.099, 0.441] & 4/11/3 & 0.5938 & 0.0156 \\
All 18 & Agentic & Single--Mean5 & 0.135 [-0.012, 0.313] & 6/11/1 & 0.1406 & 0.0156 \\
New 6 & Bounded & Max3--Mean3 & -0.792 [-1.666, -0.065] & 0/2/4 & 0.1250 & 0.1250 \\
New 6 & Bounded & Single--Mean5 & -0.233 [-1.347, 0.510] & 3/1/2 & 0.9375 & 0.0625 \\
New 6 & Agentic & Max3--Mean3 & 0.297 [-0.372, 1.264] & 1/4/1 & 1.0000 & 0.5000 \\
New 6 & Agentic & Single--Mean5 & 0.332 [0.000, 0.715] & 2/4/0 & 0.5000 & 0.5000 \\
\bottomrule
\end{tabular}
\end{table}

\begin{table}[ht]
\centering\small
\caption{Initial prospective ITT contrasts. The original point estimates and exact tests are unchanged; intervals here use the bootstrap seed 20261003. The observed nonsignificant values accompany sparse task support.}
\label{tab:initial-support}
\begin{tabular}{lllrrr}
\toprule
\rowcolor{ideablue!8} Process & Contrast & Difference [95\% CI] & P/T/N & Raw $p$ & Min. $p$ \\
\midrule
Bounded & Max3--Mean3 & 0.354 [-0.098, 1.027] & 2/8/2 & 0.500 & 0.125 \\
Bounded & Max5--Mean5 & 0.387 [-0.046, 1.027] & 2/9/1 & 0.500 & 0.250 \\
Bounded & Single--Mean3 & -0.033 [-0.098, 0.000] & 0/11/1 & 1.000 & 1.000 \\
Bounded & Single--Mean5 & -0.033 [-0.098, 0.000] & 0/11/1 & 1.000 & 1.000 \\
Agentic & Max3--Mean3 & 0.206 [0.000, 0.617] & 1/11/0 & 1.000 & 1.000 \\
Agentic & Max5--Mean5 & 0.206 [0.000, 0.617] & 1/11/0 & 1.000 & 1.000 \\
Agentic & Single--Mean3 & 0.000 [0.000, 0.000] & 0/12/0 & 1.000 & 1.000 \\
Agentic & Single--Mean5 & 0.000 [0.000, 0.000] & 0/12/0 & 1.000 & 1.000 \\
\bottomrule
\end{tabular}
\end{table}

\begin{table}[ht]
\centering\small
\caption{Leave-one-task-out ranges of mean empirical-reference loss for every follow-up policy. The task weights are recomputed equally over the remaining seventeen tasks.}
\label{tab:leaveout}
\begin{tabular}{llrlr}
\toprule
\rowcolor{ideablue!8} Process & Policy & Full mean & Leave-one-task-out range & Zero-loss tasks \\
\midrule
Bounded & Single & 0.485 & [0.370, 0.514] & 7 \\
Bounded & Mean3 & 0.472 & [0.321, 0.499] & 8 \\
Bounded & Max3 & 0.330 & [0.201, 0.349] & 11 \\
Bounded & Mean5 & 0.485 & [0.200, 0.514] & 12 \\
Bounded & Max5 & 0.303 & [0.172, 0.320] & 14 \\
Agentic & Single & 0.408 & [0.234, 0.432] & 10 \\
Agentic & Mean3 & 0.320 & [0.116, 0.339] & 11 \\
Agentic & Max3 & 0.436 & [0.237, 0.462] & 13 \\
Agentic & Mean5 & 0.273 & [0.064, 0.289] & 15 \\
Agentic & Max5 & 0.476 & [0.280, 0.504] & 13 \\
\bottomrule
\end{tabular}
\end{table}

\begin{table}[ht]
\centering\small
\caption{Post-outcome baseline-headroom sensitivity. The 98.8\% baseline threshold selects six high-baseline and twelve lower-baseline tasks.}
\label{tab:headroom}
\begin{tabular}{llllr}
\toprule
\rowcolor{ideablue!8} Process & Baseline group & Contrast & Difference [95\% CI] & Raw $p$ \\
\midrule
Bounded & At least 98.8\% & Max3--Mean3 & 0.146 [-0.023, 0.461] & 1.000 \\
Bounded & At least 98.8\% & Single--Mean5 & 0.098 [-0.013, 0.307] & 1.000 \\
Bounded & Below 98.8\% & Max3--Mean3 & -0.286 [-0.917, 0.350] & 0.406 \\
Bounded & Below 98.8\% & Single--Mean5 & -0.049 [-0.677, 0.441] & 0.883 \\
Agentic & At least 98.8\% & Max3--Mean3 & 0.033 [0.000, 0.098] & 1.000 \\
Agentic & At least 98.8\% & Single--Mean5 & 0.022 [0.000, 0.066] & 1.000 \\
Agentic & Below 98.8\% & Max3--Mean3 & 0.158 [-0.164, 0.644] & 0.688 \\
Agentic & Below 98.8\% & Single--Mean5 & 0.192 [-0.028, 0.448] & 0.156 \\
\bottomrule
\end{tabular}
\end{table}

\subsection{Penalty sensitivity}
Failure events are identified from stored execution and seed records. A numerically equal successful utility is retained as an observed outcome. Frozen-choice sensitivity changes reference utilities while preserving the original policy distributions. Reselection sensitivity recomputes every subset and card choice from the same development bank under the changed penalties. No new programs or outcomes enter either calculation. Partial fidelity has zero penalty throughout this grid; fidelity-invalid labels have penalty 0.10. The release gives all twelve failure/drift combinations, five policies, both selection modes, and every task.

\begin{table}[ht]
\centering\small
\caption{Failure-utility sensitivity at drift penalty 0.02. Columns give Max3--Mean3 and Single--Mean5 differences under frozen choices and recomputed choices. Entries average ITT utility points over the eighteen tasks.}
\label{tab:penalty}
\begin{tabular}{llrrrr}
\toprule
\rowcolor{ideablue!8} Failure utility & Process & Frozen M3 contrast & Frozen S--M5 & Reselected M3 & Reselected S--M5 \\
\midrule
0.00 & Bounded & -0.203 & -0.065 & -0.100 & 0.034 \\
0.00 & Agentic & 0.115 & 0.151 & 0.115 & 0.151 \\
-0.05 & Bounded & -0.173 & -0.032 & -0.090 & -0.032 \\
-0.05 & Agentic & 0.115 & 0.143 & 0.115 & 0.143 \\
-0.10 & Bounded & -0.142 & 0.000 & -0.142 & 0.000 \\
-0.10 & Agentic & 0.116 & 0.135 & 0.116 & 0.135 \\
-0.20 & Bounded & -0.080 & 0.065 & -0.080 & -0.059 \\
-0.20 & Agentic & 0.118 & 0.120 & 0.118 & 0.120 \\
\bottomrule
\end{tabular}
\end{table}

\subsection{Cross-split association and retained-code decomposition}
At least three paired successful model-judged faithful programs are required per card. Development scores use seed zero; final scores are compared using both seed zero and the mean of three seeds. Within-card sample covariance matrices are averaged over cards within each task and then over tasks. Correlations are computed from these pooled covariances. Bootstrap resampling acts on tasks and repeats that calculation. This weights task covariance contributions equally; correlation still depends on each task's variance scale. The output preserves all paired scores and centered deviations.

\begin{table}[ht]
\centering\small
\caption{Within-card program correlations across evaluation splits. The final two columns use seed-zero and three-seed-mean final scores, respectively.}
\label{tab:transport}
\begin{tabular}{llrrll}
\toprule
\rowcolor{ideablue!8} Process & Task group & Cards & Programs & Final seed zero & Final three-seed mean \\
\midrule
Bounded & Continuing 12 & 48 & 237 & 0.590 [0.290, 0.747] & 0.657 [0.355, 0.791] \\
Bounded & New 6 & 22 & 108 & 0.595 [0.498, 0.939] & 0.641 [0.511, 0.925] \\
Bounded & All 18 & 70 & 345 & 0.592 [0.444, 0.732] & 0.648 [0.494, 0.793] \\
Agentic & Continuing 12 & 48 & 240 & 0.656 [0.024, 0.864] & 0.681 [-0.030, 0.873] \\
Agentic & New 6 & 22 & 110 & 0.754 [0.743, 0.935] & 0.799 [0.744, 0.935] \\
Agentic & All 18 & 70 & 350 & 0.676 [0.359, 0.830] & 0.715 [0.356, 0.846] \\
\bottomrule
\end{tabular}
\end{table}

\begin{table}[ht]
\centering\small
\caption{Mean5 retained-code decomposition in ITT utility points. The first difference equals the sum of the next two mean components. Nonretained means the four complementary selection-bank programs of the same selected card, with frozen world weights.}
\label{tab:retention-decomposition}
\begin{tabular}{llllll}
\toprule
\rowcolor{ideablue!8} Process & Group & Retained--fresh & Within bank & Bank--fresh & Nonretained--fresh \\
\midrule
Bounded & Continuing 12 & 0.42 [0.04, 1.03] & 0.31 [0.00, 0.87] & 0.11 [0.01, 0.26] & 0.03 [-0.14, 0.23] \\
Bounded & New 6 & 0.29 [0.02, 0.70] & 0.57 [0.15, 1.24] & -0.28 [-1.14, 0.48] & -0.43 [-1.44, 0.43] \\
Bounded & All 18 & 0.38 [0.08, 0.82] & 0.40 [0.08, 0.84] & -0.02 [-0.33, 0.24] & -0.12 [-0.49, 0.18] \\
Agentic & Continuing 12 & 0.15 [0.00, 0.36] & 0.12 [0.01, 0.32] & 0.03 [-0.02, 0.12] & 0.00 [-0.09, 0.11] \\
Agentic & New 6 & 1.04 [0.06, 2.92] & 0.82 [0.04, 2.14] & 0.22 [-0.21, 0.82] & 0.02 [-0.29, 0.31] \\
Agentic & All 18 & 0.45 [0.05, 1.13] & 0.35 [0.04, 0.85] & 0.10 [-0.05, 0.31] & 0.01 [-0.11, 0.12] \\
\bottomrule
\end{tabular}
\end{table}

The twelve continuing tasks compare earlier selection banks with later reference banks. The six new tasks place both banks in the new collection campaign. The selected-bank contrast includes finite-bank variation and the selection weights, as shown by $H-A/4$ in Appendix~\ref{app:proofs}. Table~\ref{tab:unweighted-banks} compares all four cards with equal weights. Task-held-out family prediction in Table~\ref{tab:retention-projection} matches the regression and empirical premium populations.

\subsection{Plug-in Gaussian calibration}
Each twelve-program reference cell contributes two six-program splits. For each fit half, the card mean averages all three seeds of all six programs. Seed variance averages their within-program sample variances. Implementation variance is the variance of three-seed program means minus one third of seed variance, bounded below by zero. The predicted score of a new program at one seed uses the sum of these components. Independent Gaussian simulation supplies full-portfolio selection probabilities, with 100,000 banks per direction shared across policies. Exact enumeration of held-out subsets supplies the observed reversal probability relative to the fit-half leader. Both directions are averaged within task. The plug-in prediction conditions on the fitted parameters, while observed agreement includes noise in the fitted ranking. Appendix~\ref{app:workflow-calibration} models both banks jointly.

\begin{table}[ht]
\centering\small
\caption{Full-portfolio Gaussian calibration on eighteen tasks. Entries are predicted reversal, observed reversal, and mean absolute task-level discrepancy, with task-bootstrap intervals. A reversal changes the selected label relative to the fitted leader.}
\label{tab:calibration}
\begin{tabular}{lllll}
\toprule
\rowcolor{ideablue!8} Process & Policy & Predicted reversal & Observed reversal & Absolute discrepancy \\
\midrule
Bounded & Single & 0.192 [0.112, 0.278] & 0.256 [0.116, 0.423] & 0.122 [0.055, 0.202] \\
Bounded & Mean3 & 0.140 [0.067, 0.220] & 0.201 [0.070, 0.362] & 0.086 [0.023, 0.170] \\
Bounded & Max3 & 0.231 [0.116, 0.367] & 0.235 [0.081, 0.418] & 0.118 [0.061, 0.186] \\
Bounded & Mean5 & 0.123 [0.055, 0.202] & 0.194 [0.059, 0.358] & 0.106 [0.039, 0.189] \\
Bounded & Max5 & 0.255 [0.124, 0.408] & 0.245 [0.083, 0.435] & 0.119 [0.059, 0.192] \\
Agentic & Single & 0.166 [0.089, 0.248] & 0.187 [0.062, 0.339] & 0.102 [0.045, 0.174] \\
Agentic & Mean3 & 0.087 [0.033, 0.152] & 0.108 [0.015, 0.240] & 0.062 [0.016, 0.135] \\
Agentic & Max3 & 0.194 [0.079, 0.332] & 0.194 [0.056, 0.364] & 0.066 [0.037, 0.099] \\
Agentic & Mean5 & 0.065 [0.018, 0.122] & 0.123 [0.000, 0.289] & 0.092 [0.010, 0.203] \\
Agentic & Max5 & 0.211 [0.081, 0.367] & 0.180 [0.025, 0.359] & 0.072 [0.035, 0.115] \\
\bottomrule
\end{tabular}
\end{table}

The two-card analytic counterpart uses the leading two fitted cards and the Gaussian gap formula. The analysis file supplies per-task predictions, observed ties, seed-only predictions, and model-implied two-draw oracle floors. The full-portfolio analysis includes each task's complete eligible candidate set, avoiding a top-two shortcut when another card has high variance. Prediction error can arise from estimating parameters with six programs, score discreteness, non-Gaussian effects, and heterogeneity. Table~\ref{tab:workflow-calibration} compares this construction with complete-procedure simulation.

\subsection{Conditional item uncertainty}
Joint correctness-pattern counts reconstruct every successful seed-zero score exactly on its original integer correct-count grid. Each task/process preserves the same test-item resample across its five selection programs and twelve reference programs per card. Missing predictions use their fixed failure utilities, and fidelity penalties use their declared values. The following intervals condition on these saved tasks, code, seed realizations, and policy weights. They accompany the primary three-seed task-bootstrap analysis as a distinct sampling calculation.

\begin{table}[ht]
\centering\small
\caption{Paired item-bootstrap uncertainty for seed-zero comparisons, with 2,000 resamples. Resampling acts on shared items within the saved tasks and programs.}
\label{tab:item-uncertainty}
\begin{tabular}{llrl}
\toprule
\rowcolor{ideablue!8} Process & Comparison & Seed-zero point & Conditional item interval \\
\midrule
Bounded & Max3--Mean3 & -0.103 & [-0.471, 0.245] \\
Bounded & Single--Mean5 & 0.031 & [-0.242, 0.289] \\
Bounded & Retained Mean5--fresh & 0.305 & [0.119, 0.489] \\
Agentic & Max3--Mean3 & 0.188 & [-0.045, 0.419] \\
Agentic & Single--Mean5 & 0.153 & [-0.033, 0.349] \\
Agentic & Retained Mean5--fresh & 0.432 & [0.194, 0.720] \\
\bottomrule
\end{tabular}
\end{table}

\begin{table}[ht]
\centering\small
\caption{Observed program dispersion and estimated item-sampling contribution, in squared accuracy points. Corrected means retain negative card-level estimates and use equal task weights. Their intervals resample tasks.}
\label{tab:dispersion}
\begin{tabular}{lrrl}
\toprule
\rowcolor{ideablue!8} Process & Observed variance & Item component & Residual dispersion [95\% CI] \\
\midrule
Bounded & 1.576 & 0.344 & 1.232 [0.178, 2.667] \\
Agentic & 1.151 & 0.366 & 0.785 [0.256, 1.578] \\
\bottomrule
\end{tabular}
\end{table}

\section{Mechanism Card and Source-Level Examples}
\label{app:cards}
The following exhibit reproduces all five fields of the archived segment interaction card used in Figure~\ref{fig:overview}. Source annotations document realized choices and their saved outcomes. They are model-assisted inspection of available code. The examples use the two model judges' frozen fidelity labels.

\begin{quote}\small
\textbf{Mechanism.} Expanding features with pairwise interaction terms allows a single HistGradientBoostingClassifier to model multiplicative relationships and more complex decision surfaces with fewer tree nodes and iterations, potentially lifting accuracy under the training budget.

\textbf{Invariants.} 1. The input to the classifier must be a transformed feature matrix that contains the original features plus a set of pairwise product features derived from them. 2. The transformation must be fitted only on the training data and then applied identically to both training and evaluation sets. 3. The classifier must be exactly one HistGradientBoostingClassifier object, trained on the transformed features, with no ensembling or additional models.

\textbf{Executor choices.} - Method for generating pairwise products (e.g., all unordered pairs, include squares, use random projections into a higher-order feature space, or use scikit-learn's PolynomialFeatures with degree=2 and interaction\_only set to either True or False). - Whether to apply a subsequent scaling step (e.g., standardization) to the expanded feature matrix. - All hyperparameters of HistGradientBoostingClassifier (learning\_rate, max\_iter, max\_leaf\_nodes, l2\_regularization, etc.) under the constraint max\_iter $\leq$ 300.

\textbf{Prediction.} Development accuracy will improve beyond the frozen baseline.

\textbf{Falsifier.} If a model built following these invariants fails to achieve higher accuracy on any valid evaluation split compared to the baseline (e.g., same test split yields accuracy $\leq$ baseline), the idea is contradicted.

\end{quote}

The card permits changing all histogram gradient-boosting hyperparameters within its iteration cap. Its measured implementation distribution therefore includes both the representation and the base learner. The displayed outcomes are accuracy changes on fold 0, while the card's prediction names development performance. Outcomes below the baseline can challenge a performance prediction while the program still follows the required intervention. We keep this distinction visible in interpreting model-fidelity labels.

\clearpage
\subsection{Annotated implementation choices}
\begin{table}[ht]
\centering\small
\caption{Annotated frozen-code examples. Each row describes observed choices; no retrained parameter-isolation experiment is implied. Source paths and exact patches accompany the run records.}
\label{tab:source-cases}
\begin{tabular}{p{.17\textwidth}p{.34\textwidth}p{.23\textwidth}p{.17\textwidth}}
\toprule
\rowcolor{ideablue!8} Program & Feature construction or kernel & Learner settings & Saved outcome \\
\midrule
Segment, I02, draw 1 & Standardize, degree-two products including squares, then rescale & HGB rate 0.08, $L_2=1$, 300 iterations & 221/231 correct, $+2.16$ pp \\
Segment, I02, draw 2 & Standardize, interaction-only products, concatenate originals again & HGB rate 0.08, $L_2=1$, 300 iterations & 217/231 correct, $+0.43$ pp \\
Segment, I02, draw 3 & Interaction-only products on raw features, then standardize & HGB rate 0.05, $L_2=0.1$, 300 iterations & 212/231 correct, $-1.73$ pp \\
Gesture, I03, slots 5 and 7 & Standardized features, RBF with \texttt{gamma=scale} & SVC $C=10$, default iteration limit & 55.87\% accuracy \\
Gesture, I03, slot 6 & Standardized features, bandwidth from mean feature variance & SVC $C=1$, 300-iteration safety cap & 17.11\% accuracy \\
Gesture, I00, follow-up slot 10 & Validation-based early stopping & Accuracy scorer with incompatible callable signature & Runtime failure, ITT utility $-0.10$ \\
\bottomrule
\end{tabular}
\end{table}

The first five rows are Bounded programs. The final runtime-error row is Agentic. The two RBF variants alter margin strength, bandwidth computation, and iteration limits together. The frozen RBF card explicitly allows these choices and prohibits internal search. Their coexistence with different accuracies gives a concrete source of realization variation, while isolating each choice's causal effect would require further intervention. The runtime example received two faithful labels before execution. It demonstrates a case in which model agreement and executable validity diverge. The release includes the controls, judgment disagreements, and each conditional result.

The gesture RBF card explicitly records the following implementation choices.

\begin{quote}\small
Choose the RBF kernel bandwidth, expressed as a positive scalar or a data-dependent scale rule. Choose the margin regularization parameter as a positive scalar. Choose the solver tolerance as a positive scalar. Optionally choose whether to use the solver’s shrinking heuristic, cache size, and maximum iteration safety limit if supported, while preserving the single-fit/no-search protocol. All such choices must instantiate the same RBF-kernel maximum-margin mechanism. The route uses the full standardized feature dimension because the matrix size is feasible for one RBF SVM fit.
\end{quote}

The mechanism requires baseline standardization and a single RBF support vector classifier trained once with uniform class weights. The card leaves margin, bandwidth, tolerance, shrinking, cache, and iteration-cap choices open. This narrower specification still admits materially different realized predictors. A low test score can challenge the performance hypothesis while the executable structure follows these invariants.

\begin{figure}[ht]
\centering
\includegraphics[width=\textwidth]{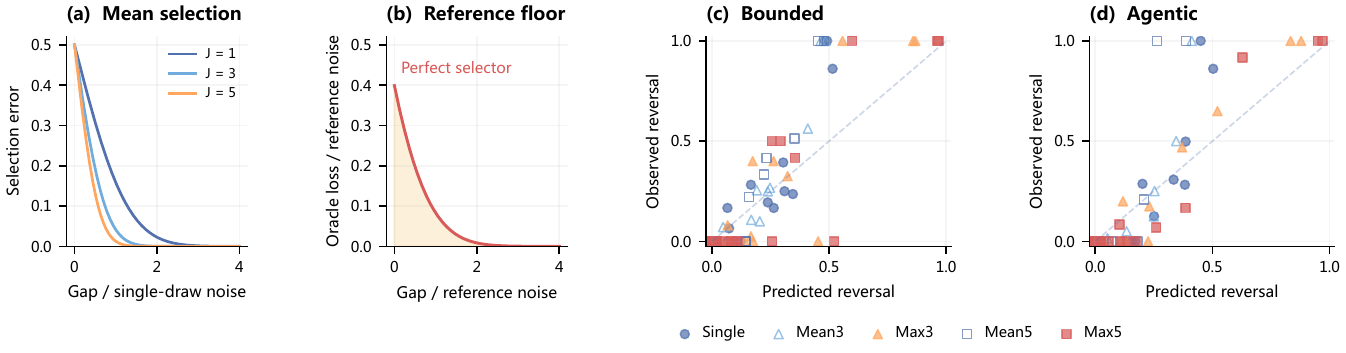}
\caption{Gaussian decision geometry and the plug-in calibration construction. The first two panels plot the two-card error and finite-reference floor under the stated model. The last two compare fixed-fit predictions with held-out agreement. Complete-procedure predictions that sample the fitted ranking appear in Figure~\ref{fig:workflow-calibration}.}
\label{fig:calibration}
\end{figure}

\clearpage
\section{Replication Decisions and Matched Model Checks}
\label{app:decision-diagnostics}
These post-outcome analyses use the frozen selection banks, shared references, source files, and cost records. Their plan fixes the primary exploratory gap threshold at nominal $\alpha=0.05$, with $0.10$ and $0.01$ sensitivities, before the replay calculations. Bootstrap resampling uses 20,000 task draws with seed 20261004. Original primary tests and their declared multiplicity family are reported separately in Appendix~\ref{app:followup-results}.

\subsection{Cost-based allocation for a card mean}
For the common nested design with $J$ implementations and $K$ seeds each, write $B=J(c_M+Kc_R)$. Substituting this budget into Equation~\ref{eq:budget} gives
\[
\Var(\widehat\Q)=\frac{(\sigma_M^2+\sigma_R^2/K)(c_M+Kc_R)}{B}.
\]
Differentiating the numerator with respect to $K>0$ gives $\sigma_M^2c_R-\sigma_R^2c_M/K^2$. Its root is Equation~\ref{eq:allocation}, the standard two-stage allocation form \citep{cochran1977sampling}. For integer designs we enumerate $K\geq1$ and set $J=\lfloor B/(c_M+Kc_R)\rfloor$, choosing the feasible pair with smallest estimated variance. A zero implementation component is handled by this same finite budget enumeration.

Cost records cover all 833 Bounded and 835 Agentic complete model-judged faithful reference programs. Generation duration comes from each assignment's recorded implementation session. Each fidelity judgment's physical request identifier links to its response duration. Seed durations come from the primary evaluation and the two saved-code reruns. The audit uses seed labels 0, 1001, and 2001. The precision-planning model assumes independent seed assignments across implementations. The common cost unit is the sum of recorded service durations. It measures work across model sessions and evaluator calls. Parallel elapsed time and monetary cost require additional scheduling and pricing inputs.

The judgment-inclusive scenario adds the two completed judgment responses to generation. The generation-only scenario isolates code-production cost. Request retries, admission, and collection overhead are outside these per-program values. The fitted common design uses the average cost and variance of the same complete-program population. Its budget equals the measured mean cost of twelve reference programs with three seeds each.

\begin{table}[!ht]
\centering\small
\caption{Recorded mean durations in seconds on the variance-analysis population. Two-judgment duration sums the primary and secondary completed responses. Seed duration averages each program's three recorded evaluations.}
\label{tab:costs}
\begin{tabular}{lrrrr}
\toprule
\rowcolor{ideablue!8} Process & Programs & Generation & Two judgments & Per seed \\
\midrule
Bounded & 833 & 21.30 & 13.95 & 13.79 \\
Agentic & 835 & 34.12 & 11.49 & 12.27 \\
\bottomrule
\end{tabular}
\end{table}

\begin{table}[!ht]
\centering\small
\caption{Common integer allocation at the measured twelve-program, three-seed reference budget. The last two columns are fitted variances in squared accuracy points for the original and proposed allocations. These are model-based planning values.}
\label{tab:allocation-cost}
\begin{tabular}{llrrlrr}
\toprule
\rowcolor{ideablue!8} Process & Program-cost definition & Budget (s) & $K^*$ & $J/K$ & Original variance & Proposed variance \\
\midrule
Bounded & Generation & 752.2 & 0.49 & 21 / 1 & 0.120 & 0.075 \\
Bounded & Generation + judgments & 919.6 & 0.63 & 18 / 1 & 0.120 & 0.088 \\
Agentic & Generation & 851.3 & 0.77 & 18 / 1 & 0.086 & 0.065 \\
Agentic & Generation + judgments & 989.2 & 0.88 & 17 / 1 & 0.086 & 0.069 \\
\bottomrule
\end{tabular}
\end{table}

\subsection{Top-two focused expansion and its gap-gated variant}
Let $z_i$ be a card's initial development score in accuracy-equivalent points. We enumerate every combination of one of the five programs for each of four cards, giving 625 initial worlds. In each world let $i_1,i_2$ be the two leading cards under the frozen tie rule. The policy expands both cards to all five programs if
\[
z_{i_1}-z_{i_2}\leq \Phi^{-1}(1-\alpha)\sqrt{\widehat v_{-s,i_1}+\widehat v_{-s,i_2}},
\qquad
\widehat v_{-s,i}=\frac{1}{17}\sum_{t\ne s}\operatorname{Var}_{j=1,\ldots,5}(D_{tij}).
\]
Variances use the other tasks' development outcomes, separately by process and mechanism family, with the declared ITT utility mapping. Each expanded card contributes its mean of five scores; the other two cards keep their initial scores. The policy then chooses among all four cards. This gives four or twelve target-task implementations. The other-task banks supply the noise calibration. All draws used by the replay come from independently generated, outcome-blind selection banks.

The Gaussian quantile supplies the nominal threshold. The replay evaluates the effects of estimating noise from other tasks and using the observed leading gap. Unconditional top-two expansion removes the gap trigger and always uses twelve programs. The matched-budget comparator expands independently of the gap with probability equal to the gap-gated policy's observed expansion frequency for that task. It therefore mixes Single and unconditional top-two expansion at the same expected implementation count. This comparator isolates the placement of expansions within the saved bank.

\begin{table}[!ht]
\centering\small
\caption{Gap-gated focused expansion on all eighteen tasks. Positive differences mean higher loss. Intervals resample paired task differences. Exact tests evaluate the exploratory contrasts; the planned nominal threshold is $\alpha=0.05$.}
\label{tab:allocation-replay}
\begin{tabular}{llrllr}
\toprule
\rowcolor{ideablue!8} Process & $\alpha$ & Programs & Gated--Mean5 & Gated--matched random & Raw $p$ \\
\midrule
Bounded & 0.10 & 11.11 & -0.045 [-0.220, 0.073] & -0.049 [-0.116, -0.006] & 0.875 \\
Bounded & 0.05 & 11.23 & -0.045 [-0.220, 0.073] & -0.035 [-0.083, -0.004] & 0.875 \\
Bounded & 0.01 & 11.50 & -0.030 [-0.177, 0.073] & -0.007 [-0.021, 0.000] & 0.875 \\
Agentic & 0.10 & 10.45 & 0.094 [0.008, 0.212] & -0.022 [-0.063, 0.007] & 0.125 \\
Agentic & 0.05 & 10.78 & 0.089 [0.008, 0.202] & -0.018 [-0.058, 0.008] & 0.125 \\
Agentic & 0.01 & 11.06 & 0.053 [0.002, 0.143] & -0.038 [-0.096, 0.000] & 0.125 \\
\bottomrule
\end{tabular}
\end{table}

\begin{table}[!ht]
\centering\small
\caption{Nominal 5\% replay by task group. The final contrast compares twelve-draw top-two expansion with twenty-draw Mean5.}
\label{tab:allocation-groups}
\begin{tabular}{llrlll}
\toprule
\rowcolor{ideablue!8} Process & Group & Gated programs & Gated loss & Gated--Mean5 & Top-two--Mean5 \\
\midrule
Bounded & Continuing 12 & 11.16 & 0.228 [0.000, 0.494] & -0.001 [-0.003, 0.000] & -0.001 [-0.003, 0.000] \\
Bounded & New 6 & 11.36 & 0.866 [0.114, 2.155] & -0.132 [-0.655, 0.222] & -0.089 [-0.528, 0.222] \\
Bounded & All 18 & 11.23 & 0.440 [0.111, 0.932] & -0.045 [-0.220, 0.073] & -0.030 [-0.177, 0.073] \\
Agentic & Continuing 12 & 10.69 & 0.417 [0.012, 1.113] & 0.020 [0.000, 0.049] & 0.010 [0.000, 0.027] \\
Agentic & New 6 & 10.94 & 0.250 [0.023, 0.507] & 0.226 [0.000, 0.484] & 0.129 [0.000, 0.386] \\
Agentic & All 18 & 10.78 & 0.361 [0.059, 0.826] & 0.089 [0.008, 0.202] & 0.049 [0.000, 0.139] \\
\bottomrule
\end{tabular}
\end{table}

Under the nominal 5\% rule, expansion occurs in 90.4\% of Bounded worlds and 84.7\% of Agentic worlds. Focusing expansion on two cards supplies most of the reduction from twenty draws. Relative to Mean5, the gap-gated contrast has four nonzero tasks in each process, with exact sign-flip $p=0.875$ and $0.125$. The Bounded gated-minus-random comparison has four negative and fourteen zero task differences, giving exact $p=0.125$ alongside its bootstrap interval. Table~\ref{tab:allocation-replay} reports both summaries so that their finite-sample resolution is visible.

\subsection{Complete-partition calibration with an estimated leader}
\label{app:workflow-calibration}
Every eligible card contains twelve complete primary model-judged faithful reference programs with three seeds. The fit-half mean averages six programs and three seeds; the held-out operator uses seed zero and a size-one, size-three, or size-five subset of the other six programs. The target is whether the held-out policy selects a different card from the estimated fit-half leader. All eighteen tasks per process have at least two eligible cards.

The complete-partition diagnostic averages all $\binom{12}{6}=924$ oriented partitions. Each common partition assigns the same six reference-slot labels to fitting in every card. Within a partition, empirical CDF products integrate every independent per-card held-out subset choice exactly. Integer correct counts determine ties before normalization. The original rule favors the largest card label; a sensitivity reverses that order. A second target asks whether the fit and held-out maximizing-card sets are disjoint, preserving all ties in both sets.

Reference-slot labels have no natural cross-card pairing. We therefore also integrate independently chosen partitions for each card. For card $i$, enumerate the empirical joint atoms $(A_i,Z_i)$ of its fit score and held-out operator across all partitions and subsets. Two-dimensional empirical CDFs give the probability that each atom wins both rankings against every other card, using strict comparisons when required by the label order. Summing these joint winner probabilities gives agreement over all independent partition combinations. CDF and point-mass products similarly integrate overlap of maximizing sets. This calculation averages the full product distribution without enumerating $924^C$ partition combinations. It is invariant to relabeling program slots separately within cards.

All partition calculations reuse the same twelve observed programs. Task-bootstrap intervals resample tasks; partition quantiles in the output describe conditional variation within the saved bank. They do not represent independent replication. Fit and selection ties, both label rules, disjoint sets, and individual task results are provided for each policy.

The Gaussian check fits means and variance components to all twelve programs, then samples 50,000 complete fit/held-out comparisons per task. A fit-half mean has variance $\sigma_M^2/6+\sigma_R^2/18$; each held-out seed-zero score has variance $\sigma_M^2+\sigma_R^2$. Independent Gaussian draws reproduce these marginals and the policy operators. Zero-variance components remain point masses. This fitted model generates no maximizing ties in these simulations, whereas empirical fit-bank ties occur in 4.4\% and 1.3\% of common partitions averaged across tasks.

\begin{table}[!ht]
\centering\small
\caption{Complete-procedure Gaussian reversal and empirical averaging over every 6/6 partition. Common partition rows align slot labels across cards; independent partitions integrate each card separately. The last column compares model and common-partition task means.}
\label{tab:workflow-calibration}
\begin{tabular}{llrrrr}
\toprule
\rowcolor{ideablue!8} Process & Policy & Gaussian & 924 common & Independent & Task MAE \\
\midrule
Bounded & Single & 0.209 & 0.255 & 0.257 & 0.127 \\
Bounded & Mean3 & 0.149 & 0.203 & 0.205 & 0.079 \\
Bounded & Max3 & 0.204 & 0.227 & 0.227 & 0.165 \\
Bounded & Mean5 & 0.128 & 0.174 & 0.174 & 0.075 \\
Bounded & Max5 & 0.213 & 0.236 & 0.237 & 0.184 \\
Agentic & Single & 0.174 & 0.191 & 0.189 & 0.101 \\
Agentic & Mean3 & 0.098 & 0.121 & 0.119 & 0.066 \\
Agentic & Max3 & 0.181 & 0.197 & 0.195 & 0.085 \\
Agentic & Mean5 & 0.076 & 0.123 & 0.118 & 0.078 \\
Agentic & Max5 & 0.192 & 0.191 & 0.189 & 0.081 \\
\bottomrule
\end{tabular}
\end{table}

\begin{table}[!ht]
\centering\small
\caption{Tie sensitivities of complete-partition empirical reversal. Disjoint sets count a reversal only when the fit and selected maximizing sets have no common card. Selection ties refer to multiple maximizers of the held-out policy score.}
\label{tab:partition-ties}
\begin{tabular}{llrrrr}
\toprule
\rowcolor{ideablue!8} Process & Policy & Largest label & Smallest label & Disjoint sets & Selection ties \\
\midrule
Bounded & Single & 0.255 & 0.195 & 0.180 & 0.049 \\
Bounded & Mean3 & 0.203 & 0.158 & 0.152 & 0.015 \\
Bounded & Max3 & 0.227 & 0.171 & 0.151 & 0.053 \\
Bounded & Mean5 & 0.174 & 0.128 & 0.122 & 0.014 \\
Bounded & Max5 & 0.236 & 0.182 & 0.156 & 0.064 \\
Agentic & Single & 0.191 & 0.184 & 0.149 & 0.072 \\
Agentic & Mean3 & 0.121 & 0.142 & 0.107 & 0.043 \\
Agentic & Max3 & 0.197 & 0.185 & 0.147 & 0.098 \\
Agentic & Mean5 & 0.123 & 0.124 & 0.111 & 0.011 \\
Agentic & Max5 & 0.191 & 0.204 & 0.151 & 0.119 \\
\bottomrule
\end{tabular}
\end{table}

For Mean5, Bounded's largest-label reversal exceeds its Gaussian prediction by 0.046; its smallest-label reversal is almost identical to that prediction. Agentic's corresponding gap is 0.046 and changes little under the opposite tie order. The disjoint-set target gives 0.122 and 0.111. Different tie targets answer different ranking questions. The remaining discrepancies also depend on finite-bank component estimates, score shape, and assumptions about program and seed dependence.

For a two-card check, each fit half supplies its two leading cards under the original label rule. The empirical statistic integrates their held-out ordering over all partitions and subsets. In the Gaussian calculation, integrating the signed population gap formula over each simulated fit pair agrees with direct held-out simulation. This numerical agreement checks the implementation of Proposition~\ref{prop:gap}; the empirical column measures its calibration on the saved bank.

\begin{table}[!ht]
\centering\small
\caption{Two-card reversal with a fitted leading pair. Formula and simulation average the same Gaussian fit-pair procedure. The empirical comparison averages all 924 common partitions.}
\label{tab:two-card-calibration}
\begin{tabular}{llrrr}
\toprule
\rowcolor{ideablue!8} Process & Policy & Integrated formula & Gaussian simulation & 924 empirical \\
\midrule
Bounded & Single & 0.183 & 0.182 & 0.228 \\
Bounded & Mean3 & 0.137 & 0.137 & 0.184 \\
Bounded & Mean5 & 0.119 & 0.119 & 0.159 \\
Agentic & Single & 0.131 & 0.130 & 0.145 \\
Agentic & Mean3 & 0.084 & 0.084 & 0.105 \\
Agentic & Mean5 & 0.070 & 0.070 & 0.107 \\
\bottomrule
\end{tabular}
\end{table}

\begin{figure}[!ht]
\centering
\includegraphics[width=.94\textwidth]{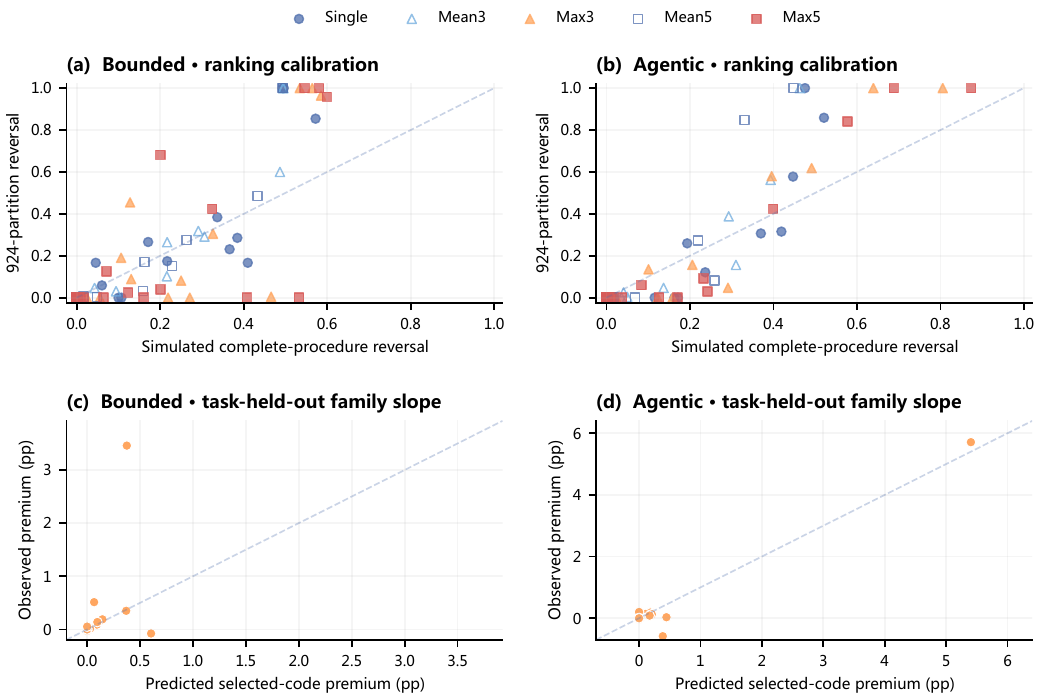}
\caption{Post-outcome model checks. Top panels compare complete-procedure Gaussian reversal with all-924-partition reversal for each task and policy under the largest-label rule. Bottom panels compare Mean5 premiums with family-pooled slopes trained on other tasks, using full-card development centering. The prediction and empirical premium use the same primary model-judged faithful complete-bank population. Dashed lines mark equal predicted and observed values.}
\label{fig:workflow-calibration}
\end{figure}

\subsection{Task-held-out family prediction of the premium}
The prediction population matches the original primary model-judged faithful complete-bank comparison: sixteen Bounded and eighteen Agentic tasks. Each retained card has five successful faithful selection programs and twelve complete faithful references. Development scores use seed zero; final scores average three seeds. Frozen selection weights operate on the original eligible card subset.

For every card in a target task, the slope is the mean within-card development--final covariance divided by the mean development variance over the same mechanism family in the other tasks. This excludes all target-task final scores from fitting and pools every program's slope. We multiply the slope by each program's development deviation and its frozen retention weight. Full-card centering uses $D_{ij}-\bar D_i$; leave-one-program-out centering multiplies each such deviation by $5/4$. Both are empirical substitutes for the population center in Equation~\ref{eq:retention}.

\begin{table}[!ht]
\centering\small
\caption{Task-held-out family slopes applied to every program. Full and leave-one-out columns change development centering while keeping the same slope. All entries are accuracy points.}
\label{tab:retention-projection}
\begin{tabular}{llrrrrr}
\toprule
\rowcolor{ideablue!8} Process & Policy & Full center & LOO center & Observed & Full MAE & LOO MAE \\
\midrule
Bounded & Single & 0.229 & 0.287 & 0.096 & 0.175 & 0.221 \\
Bounded & Mean3 & 0.162 & 0.203 & 0.161 & 0.142 & 0.145 \\
Bounded & Max3 & 0.465 & 0.581 & 0.216 & 0.325 & 0.426 \\
Bounded & Mean5 & 0.111 & 0.139 & 0.294 & 0.276 & 0.283 \\
Bounded & Max5 & 0.527 & 0.659 & 0.223 & 0.421 & 0.532 \\
Agentic & Single & 0.188 & 0.235 & 0.156 & 0.238 & 0.239 \\
Agentic & Mean3 & 0.292 & 0.364 & 0.356 & 0.125 & 0.097 \\
Agentic & Max3 & 0.483 & 0.604 & 0.490 & 0.347 & 0.343 \\
Agentic & Mean5 & 0.386 & 0.482 & 0.328 & 0.120 & 0.178 \\
Agentic & Max5 & 0.690 & 0.863 & 0.733 & 0.449 & 0.539 \\
\bottomrule
\end{tabular}
\end{table}

Mean5's full-center predictions are 0.111 and 0.386 against observed premiums of 0.294 and 0.328, with task MAEs 0.276 and 0.120. Leave-one-program-out centering gives predictions 0.139 and 0.482, with MAEs 0.283 and 0.178. Pooling slopes improves Agentic prediction relative to the original four-program card fits; Bounded task discrepancies remain as large as its mean premium. The observed cross-split association supports a separate descriptive finding from this predictive approximation.

\begin{figure}[!ht]
\centering
\includegraphics[width=.94\textwidth]{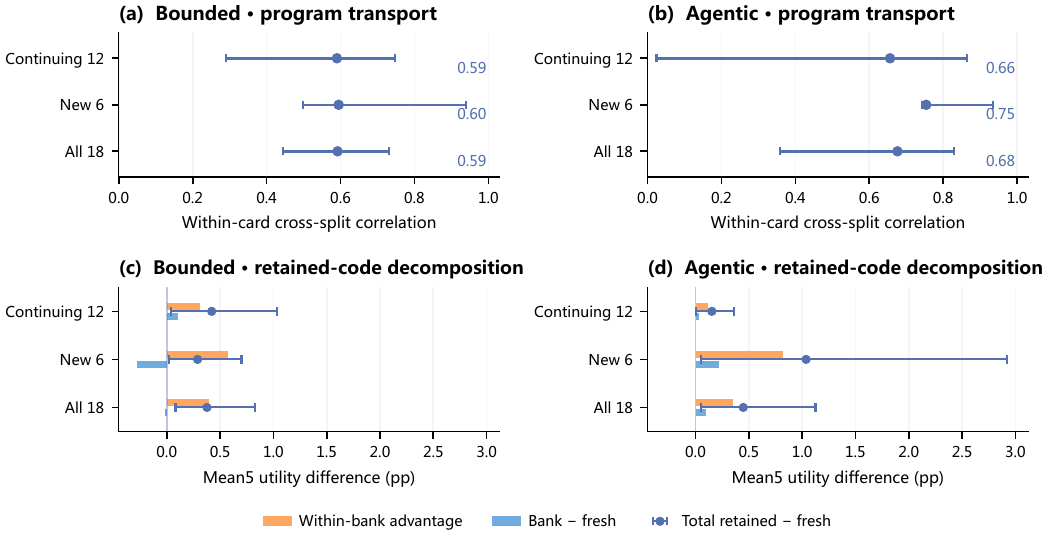}
\caption{Post-outcome transport and bank decomposition. Top panels use task-weighted covariance matrices from paired model-judged faithful programs. Bottom panels decompose the ITT Mean5 premium into within-bank selected-program advantage and the selected-card bank--fresh gap. Points and intervals summarize total premiums; bars give mean components. Table~\ref{tab:premium-main} reports their task concentration.}
\label{fig:transport}
\end{figure}

\subsection{Task influence, ranks, and bank comparisons}
For each task, the covariance analysis averages the within-card covariance matrices of paired program scores. It then averages task covariances and computes their correlation. This gives equal task weights at the covariance stage. The final ratio reflects each task's and card's variance scale. We report leave-one-task-out ranges and median within-card rank correlations to expose that weighting.

Spearman correlations use average ranks for tied scores and require positive score variance on both splits. Forty-one of seventy paired cards satisfy this requirement in each process. A second sensitivity removes any paired program whose development or final score differs by more than ten accuracy points from its within-card median, then recomputes covariances on cards with at least three remaining programs. It removes zero Bounded programs and one Agentic program. The low-accuracy gesture RBF example occupies initial reference slot 6; the transport calculation uses selection slots 0--4.

\begin{table}[!ht]
\centering\small
\caption{Transport under task deletion, rank correlation, and the fixed ten-point outlier rule. Rank medians include cards with variation on both splits; the support column gives informative rank cards over all paired cards.}
\label{tab:transport-robustness}
\begin{tabular}{llrlrlr}
\toprule
\rowcolor{ideablue!8} Process & Group & Covariance $r$ & Task-deletion range & Median rank $r$ & Support & Trimmed $r$ \\
\midrule
Bounded & Continuing 12 & 0.590 & [0.491, 0.651] & 0.829 & 22/48 & 0.590 \\
Bounded & New 6 & 0.595 & [0.524, 0.770] & 0.667 & 19/22 & 0.595 \\
Bounded & All 18 & 0.592 & [0.552, 0.630] & 0.745 & 41/70 & 0.592 \\
Agentic & Continuing 12 & 0.656 & [0.411, 0.737] & 0.229 & 27/48 & 0.484 \\
Agentic & New 6 & 0.754 & [0.749, 0.771] & 1.000 & 14/22 & 0.754 \\
Agentic & All 18 & 0.676 & [0.570, 0.738] & 0.791 & 41/70 & 0.590 \\
\bottomrule
\end{tabular}
\end{table}

The equal-card bank statistic is $\frac14\sum_{i=1}^4(\bar T_i-q_i)$. It averages every card with weight one quarter, independent of the selected-card probabilities. ITT uses all four cards. The model-judged successful sensitivity averages cards with at least one complete faithful program in each bank. It covers three Bounded and two Agentic cards on isolet and all four cards on every other task. Both seed-zero and three-seed targets are reported. The selected-card statistic in Table~\ref{tab:retention-decomposition} is a separate quantity. For example, the continuing Bounded selected-card bank gap is about 0.11 points, whereas the equal-card gap is $-0.02$.

\begin{table}[!ht]
\centering\small
\caption{Selection-bank minus fresh-reference utility in accuracy-equivalent points. ITT weights all four cards equally. The successful-program sensitivity weights its available cards equally and uses unpenalized accuracy. Task groups and seed targets are shown separately.}
\label{tab:unweighted-banks}
\begin{tabular}{lllll}
\toprule
\rowcolor{ideablue!8} Process & Group & ITT, three seeds & Successful, three seeds & ITT, seed zero \\
\midrule
Bounded & Continuing 12 & -0.022 [-0.197, 0.131] & -0.032 [-0.164, 0.064] & -0.006 [-0.188, 0.152] \\
Bounded & New 6 & -0.419 [-1.041, 0.099] & -0.290 [-0.736, 0.034] & -0.374 [-1.023, 0.087] \\
Bounded & All 18 & -0.154 [-0.414, 0.055] & -0.118 [-0.300, 0.020] & -0.129 [-0.392, 0.075] \\
Agentic & Continuing 12 & -0.047 [-0.129, 0.030] & -0.069 [-0.174, 0.024] & -0.081 [-0.201, 0.021] \\
Agentic & New 6 & -0.019 [-0.418, 0.330] & -0.140 [-0.473, 0.075] & -0.004 [-0.419, 0.341] \\
Agentic & All 18 & -0.038 [-0.180, 0.091] & -0.093 [-0.222, 0.010] & -0.055 [-0.212, 0.088] \\
\bottomrule
\end{tabular}
\end{table}

Both banks use official final fold 0, training folds 1--9, the same baseline and evaluator contract, and single-threaded CPU fits. The continuing selection bank's final scores come from the initial collection; its fresh reference scores come from the later collection. The campaigns have different evaluation schedules. The recorded runtime diagnostic evaluates the complete banks of timeout-affected tasks at a common load, and Appendix~\ref{app:runtime-results} gives that comparison. Differences between banks can reflect finite program draws, collection batches, and operational failures; the equal-card statistic directly measures their combined mean shift.

\subsection{Selected-code gains as idea estimates}
For Mean5, let $\bar G$ be the task-average selected-code premium and $\bar A$ the task-average selected-code gain over baseline on final data. We report $\bar G/\bar A$, the ratio of these aggregate means. This differs from averaging task ratios, which can be undefined at zero gain. Both numerator and denominator use the same frozen selection weights and three-seed final outcomes.

\begin{table}[!ht]
\centering\small
\caption{Mean5 selected-code gain, its premium over fresh implementation, and their point ratio. The final column is the fraction of task-bootstrap draws with nonpositive mean selected-code gain. The task-bootstrap seed is 20261004 for these intervals.}
\label{tab:optimism-ratio}
\begin{tabular}{llllrr}
\toprule
\rowcolor{ideablue!8} Process & Population & Selected-code gain & Premium over fresh & Point ratio & Nonpositive gain \\
\midrule
Bounded & ITT & 0.481 [-0.005, 0.984] & 0.377 [0.082, 0.813] & 78.4\% & 2.6\% \\
Bounded & Model-judged & 0.334 [-0.188, 0.893] & 0.294 [0.029, 0.753] & 87.9\% & 11.1\% \\
Agentic & ITT & 0.802 [-0.130, 1.963] & 0.448 [0.047, 1.128] & 55.9\% & 5.0\% \\
Agentic & Model-judged & 0.856 [-0.046, 1.985] & 0.328 [-0.048, 0.997] & 38.3\% & 3.3\% \\
\bottomrule
\end{tabular}
\end{table}

The ITT point ratios are 78\% and 56\%. The denominator intervals cross zero, so the absolute paired premium supplies the primary uncertainty summary for this comparison. Machine-readable outputs also include the signed bootstrap ratio distribution. The premium measures the error incurred by using the selected-code result as the mean of the sampled fresh implementation process.

\subsection{Static source strata and paired correctness}
The source-only parser reads constructor literals and import aliases. It compares learning rate, iterations, maximum leaves, minimum leaf size, and $L_2$ regularization with the baseline. Dynamic arguments and parameter mutations receive separate tags. RBF constructors identify the alternative predictor family. These tags accompany the frozen model-fidelity judgments.

\begin{table}[!ht]
\centering\small
\caption{Source tags for all 3,048 assignment records. Baseline and changed columns refer to the five checked gradient-boosting settings. The final column covers other constructors and dynamic call structure.}
\label{tab:source-strata-counts}
\begin{tabular}{llrrrrr}
\toprule
\rowcolor{ideablue!8} Cohort & Process & Baseline core & Changed core & Dynamic core & RBF & Other \\
\midrule
Archived & Bounded & 89 & 32 & 3 & 6 & 26 \\
Archived & Agentic & 81 & 40 & 3 & 5 & 27 \\
Initial & Bounded & 288 & 0 & 0 & 95 & 1 \\
Initial & Agentic & 282 & 0 & 4 & 96 & 2 \\
Follow-up & Bounded & 737 & 1 & 0 & 243 & 3 \\
Follow-up & Agentic & 725 & 0 & 12 & 246 & 1 \\
\bottomrule
\end{tabular}
\end{table}

\begin{table}[!ht]
\centering\small
\caption{Follow-up model-judged faithful reference variation within source strata. Cards contribute at least two programs. Implementation variance subtracts the estimated within-program seed component from pooled within-card score variance. All displayed strata span eighteen tasks.}
\label{tab:source-strata-variance}
\begin{tabular}{llrrlr}
\toprule
\rowcolor{ideablue!8} Process & Source stratum & Programs & Cards & Implementation SD (pp) & Seed SD (pp) \\
\midrule
Bounded & Baseline HGB core & 616 & 53 & 1.284 [0.413, 1.893] & 0.530 \\
Bounded & RBF family & 214 & 18 & 0.757 [0.390, 1.049] & 0.000 \\
Agentic & Baseline HGB core & 609 & 52 & 1.045 [0.552, 1.538] & 0.510 \\
Agentic & RBF family & 216 & 18 & 0.840 [0.361, 1.195] & 0.000 \\
\bottomrule
\end{tabular}
\end{table}

Archived cards allow learner-setting changes, including the segment card. Prospective stopping and feature cards freeze the five checked settings. The sole follow-up change, Bounded isolet I00 slot 7, raises iterations from 150 to 300. Both judges label it partial; ITT includes its outcome and the faithful variance analysis excludes it. Each stratum uses its own card and program support.

Recognized support-vector sources contain two archived, one initial, and four follow-up literal iteration caps, plus one dynamic follow-up cap. The controller saves successful scores and predictions, with captured standard error for failures. The original successful runs lack warning and fitted-iteration records. Separate instrumented reruns reproduce all eight saved prediction vectors and confirm that all seven literal caps bind (Table~\ref{tab:svc-convergence}).

For the segment example, saved joint correctness counts give 211 items correct under both draws 1 and 3, ten correct only under draw 1, one correct only under draw 3, and nine incorrect under both. The exact paired test conditions on the eleven discordant items and evaluates a two-sided binomial probability under equal success probabilities \citep{mcnemar1947paired}. The paired item-bootstrap resamples the same 231 items for both frozen programs, giving the 3.90-point difference and [1.30, 6.93] interval. This archived post-outcome illustration permits learner-setting changes; Figure~\ref{fig:overview} uses the median-rule follow-up card with baseline core settings. The supplement includes these joint counts and all three programs' correctness-pattern counts.

\section{Premium Robustness and Complete-Partition Diagnostics}
\label{app:premium-robustness}
The selected-code endpoint is a prespecified secondary comparison. The local analysis plan was initialized at 10:34:49 UTC on October 2, 2026 and frozen at 10:51:03, explicitly naming the highest-selection-score program versus fresh realizations on the final split with three seeds. All new model calls ended by 11:50:27. Development evaluation fixed policy weights at 12:28:58; new final evaluation began at 12:30:11. The original continuing-task fold-0 outcomes had already been observed. The robustness, complete-partition, and rule-selected example analyses were added after outcomes and use seed 20261005 for 20,000 task or paired-item bootstrap draws. Saved endpoint intervals retain the original seed 20261002.

\subsection{Premium task influence across policies and populations}
The tables report all five policies in each population and task group. Exact sign flips use the absolute mean statistic, setting magnitudes at most $10^{-12}$ to zero. With $m$ nonzero tasks, their minimum possible two-sided probability is $\min(1,2^{1-m})$. The JSON contains this resolution, each task's contribution, every deletion mean, and the largest task's share of the signed sum. Deletion ranges measure point-estimate influence. Each deleted sample also receives a bootstrap interval for pooled Single and Mean5.

\begin{table}[!ht]
\centering\small
\caption{Bounded ITT selected-code premiums. Task-deletion ranges contain means; probabilities are exact unadjusted sign flips.}
\label{tab:premium-lite-itt}
\begin{tabular}{llrlrclr}
\toprule
\rowcolor{ideablue!8} Group & Policy & Tasks & Mean [95\% interval] & Median & P/T/N & Deletion means & $p$ \\
\midrule
Continuing & Single & 12 & 0.273 [0.093, 0.474] & 0.052 & 7/4/1 & [0.217, 0.301] & 0.016 \\
Continuing & Mean3 & 12 & 0.274 [0.060, 0.548] & 0.071 & 7/4/1 & [0.170, 0.302] & 0.016 \\
Continuing & Max3 & 12 & 0.404 [0.155, 0.671] & 0.121 & 7/4/1 & [0.327, 0.443] & 0.016 \\
Continuing & Mean5 & 12 & 0.421 [0.038, 1.034] & 0.000 & 5/6/1 & [0.145, 0.462] & 0.062 \\
Continuing & Max5 & 12 & 0.369 [0.134, 0.626] & 0.121 & 7/4/1 & [0.289, 0.406] & 0.016 \\
New six & Single & 6 & 0.258 [-0.023, 0.646] & 0.159 & 4/0/2 & [0.078, 0.342] & 0.219 \\
New six & Mean3 & 6 & 0.259 [-0.015, 0.670] & 0.108 & 4/0/2 & [0.064, 0.338] & 0.188 \\
New six & Max3 & 6 & 0.362 [0.071, 0.738] & 0.244 & 5/0/1 & [0.188, 0.451] & 0.094 \\
New six & Mean5 & 6 & 0.289 [0.017, 0.686] & 0.107 & 5/0/1 & [0.099, 0.362] & 0.125 \\
New six & Max5 & 6 & 0.353 [0.051, 0.730] & 0.244 & 5/0/1 & [0.176, 0.452] & 0.125 \\
Pooled & Single & 18 & 0.268 [0.109, 0.453] & 0.052 & 11/4/3 & [0.216, 0.294] & 0.005 \\
Pooled & Mean3 & 18 & 0.269 [0.087, 0.495] & 0.084 & 11/4/3 & [0.201, 0.293] & 0.004 \\
Pooled & Max3 & 18 & 0.390 [0.192, 0.608] & 0.164 & 12/4/2 & [0.339, 0.418] & 0.001 \\
Pooled & Mean5 & 18 & 0.377 [0.082, 0.820] & 0.031 & 10/6/2 & [0.196, 0.404] & 0.007 \\
Pooled & Max5 & 18 & 0.364 [0.174, 0.578] & 0.164 & 12/4/2 & [0.312, 0.394] & 0.001 \\
\bottomrule
\end{tabular}
\end{table}

\begin{table}[!ht]
\centering\small
\caption{Bounded model-judged faithful selected-code premiums. Task-deletion ranges contain means; probabilities are exact unadjusted sign flips.}
\label{tab:premium-lite-faithful}
\begin{tabular}{llrlrclr}
\toprule
\rowcolor{ideablue!8} Group & Policy & Tasks & Mean [95\% interval] & Median & P/T/N & Deletion means & $p$ \\
\midrule
Continuing & Single & 11 & 0.112 [0.016, 0.230] & 0.000 & 5/5/1 & [0.071, 0.127] & 0.062 \\
Continuing & Mean3 & 11 & 0.198 [0.009, 0.486] & 0.000 & 4/6/1 & [0.075, 0.220] & 0.125 \\
Continuing & Max3 & 11 & 0.230 [0.049, 0.448] & 0.000 & 5/5/1 & [0.161, 0.256] & 0.062 \\
Continuing & Mean5 & 11 & 0.380 [0.009, 1.036] & 0.000 & 4/6/1 & [0.072, 0.421] & 0.125 \\
Continuing & Max5 & 11 & 0.237 [0.041, 0.488] & 0.000 & 5/5/1 & [0.141, 0.264] & 0.062 \\
New six & Single & 5 & 0.060 [-0.039, 0.173] & 0.068 & 3/0/2 & [0.005, 0.096] & 0.438 \\
New six & Mean3 & 5 & 0.082 [0.037, 0.119] & 0.091 & 4/0/1 & [0.068, 0.103] & 0.125 \\
New six & Max3 & 5 & 0.185 [0.044, 0.333] & 0.139 & 4/0/1 & [0.132, 0.252] & 0.125 \\
New six & Mean5 & 5 & 0.103 [-0.010, 0.230] & 0.091 & 4/0/1 & [0.041, 0.148] & 0.188 \\
New six & Max5 & 5 & 0.192 [0.010, 0.355] & 0.152 & 4/0/1 & [0.124, 0.276] & 0.188 \\
Pooled & Single & 16 & 0.096 [0.022, 0.184] & 0.026 & 8/5/3 & [0.067, 0.108] & 0.035 \\
Pooled & Mean3 & 16 & 0.161 [0.033, 0.361] & 0.026 & 8/6/2 & [0.077, 0.174] & 0.008 \\
Pooled & Max3 & 16 & 0.216 [0.080, 0.371] & 0.088 & 9/5/2 & [0.169, 0.236] & 0.006 \\
Pooled & Mean5 & 16 & 0.294 [0.030, 0.749] & 0.005 & 8/6/2 & [0.083, 0.318] & 0.025 \\
Pooled & Max5 & 16 & 0.223 [0.075, 0.406] & 0.095 & 9/5/2 & [0.158, 0.247] & 0.009 \\
\bottomrule
\end{tabular}
\end{table}

\begin{table}[!ht]
\centering\small
\caption{Agentic ITT selected-code premiums. Task-deletion ranges contain means; probabilities are exact unadjusted sign flips.}
\label{tab:premium-agent_core-itt}
\begin{tabular}{llrlrclr}
\toprule
\rowcolor{ideablue!8} Group & Policy & Tasks & Mean [95\% interval] & Median & P/T/N & Deletion means & $p$ \\
\midrule
Continuing & Single & 12 & 0.070 [-0.047, 0.214] & 0.000 & 5/5/2 & [0.013, 0.106] & 0.281 \\
Continuing & Mean3 & 12 & 0.148 [0.008, 0.344] & 0.000 & 5/5/2 & [0.072, 0.163] & 0.094 \\
Continuing & Max3 & 12 & 0.083 [-0.154, 0.325] & 0.007 & 6/5/1 & [0.001, 0.174] & 0.500 \\
Continuing & Mean5 & 12 & 0.153 [0.004, 0.362] & 0.000 & 5/5/2 & [0.073, 0.169] & 0.125 \\
Continuing & Max5 & 12 & 0.038 [-0.309, 0.332] & 0.007 & 6/5/1 & [-0.053, 0.178] & 0.766 \\
New six & Single & 6 & 0.504 [0.021, 1.355] & 0.101 & 3/1/2 & [0.095, 0.613] & 0.250 \\
New six & Mean3 & 6 & 0.749 [0.057, 2.057] & 0.134 & 5/1/0 & [0.103, 0.899] & 0.062 \\
New six & Max3 & 6 & 1.048 [0.086, 2.423] & 0.192 & 5/1/0 & [0.411, 1.257] & 0.062 \\
New six & Mean5 & 6 & 1.037 [0.057, 2.923] & 0.134 & 5/1/0 & [0.103, 1.245] & 0.062 \\
New six & Max5 & 6 & 1.472 [0.086, 3.341] & 0.192 & 5/1/0 & [0.624, 1.766] & 0.062 \\
Pooled & Single & 18 & 0.215 [0.010, 0.532] & 0.000 & 8/6/4 & [0.078, 0.247] & 0.080 \\
Pooled & Mean3 & 18 & 0.348 [0.052, 0.828] & 0.029 & 10/6/2 & [0.135, 0.370] & 0.003 \\
Pooled & Max3 & 18 & 0.405 [0.015, 0.952] & 0.038 & 11/6/1 & [0.180, 0.482] & 0.082 \\
Pooled & Mean5 & 18 & 0.448 [0.050, 1.124] & 0.024 & 10/6/2 & [0.138, 0.476] & 0.004 \\
Pooled & Max5 & 18 & 0.516 [-0.037, 1.273] & 0.038 & 11/6/1 & [0.211, 0.635] & 0.173 \\
\bottomrule
\end{tabular}
\end{table}

\begin{table}[!ht]
\centering\small
\caption{Agentic model-judged faithful selected-code premiums. Task-deletion ranges contain means; probabilities are exact unadjusted sign flips.}
\label{tab:premium-agent_core-faithful}
\begin{tabular}{llrlrclr}
\toprule
\rowcolor{ideablue!8} Group & Policy & Tasks & Mean [95\% interval] & Median & P/T/N & Deletion means & $p$ \\
\midrule
Continuing & Single & 12 & -0.019 [-0.195, 0.157] & 0.000 & 4/5/3 & [-0.085, 0.045] & 0.938 \\
Continuing & Mean3 & 12 & 0.159 [0.006, 0.433] & 0.000 & 5/5/2 & [0.028, 0.176] & 0.094 \\
Continuing & Max3 & 12 & 0.211 [-0.191, 0.794] & 0.007 & 6/5/1 & [-0.047, 0.313] & 0.516 \\
Continuing & Mean5 & 12 & -0.026 [-0.139, 0.047] & 0.000 & 4/5/3 & [-0.041, 0.024] & 0.891 \\
Continuing & Max5 & 12 & 0.364 [-0.339, 1.400] & 0.007 & 6/5/1 & [-0.101, 0.533] & 0.516 \\
New six & Single & 6 & 0.504 [0.021, 1.355] & 0.101 & 3/1/2 & [0.095, 0.613] & 0.250 \\
New six & Mean3 & 6 & 0.749 [0.057, 2.057] & 0.134 & 5/1/0 & [0.103, 0.899] & 0.062 \\
New six & Max3 & 6 & 1.048 [0.086, 2.423] & 0.192 & 5/1/0 & [0.411, 1.257] & 0.062 \\
New six & Mean5 & 6 & 1.037 [0.057, 2.923] & 0.134 & 5/1/0 & [0.103, 1.245] & 0.062 \\
New six & Max5 & 6 & 1.472 [0.086, 3.341] & 0.192 & 5/1/0 & [0.624, 1.766] & 0.062 \\
Pooled & Single & 18 & 0.156 [-0.079, 0.489] & 0.000 & 7/6/5 & [0.015, 0.207] & 0.414 \\
Pooled & Mean3 & 18 & 0.356 [0.038, 0.871] & 0.029 & 10/6/2 & [0.142, 0.378] & 0.003 \\
Pooled & Max3 & 18 & 0.490 [0.008, 1.110] & 0.038 & 11/6/1 & [0.270, 0.572] & 0.125 \\
Pooled & Mean5 & 18 & 0.328 [-0.049, 0.996] & 0.007 & 9/6/3 & [0.012, 0.382] & 0.431 \\
Pooled & Max5 & 18 & 0.733 [-0.024, 1.667] & 0.038 & 11/6/1 & [0.440, 0.864] & 0.125 \\
\bottomrule
\end{tabular}
\end{table}

\begin{table}[!ht]
\centering\small
\caption{Largest signed contribution and task-deletion interval sensitivity. The last column is the smallest lower bootstrap endpoint across every deleted task, in points.}
\label{tab:premium-influence}
\begin{tabular}{lllr rrr}
\toprule
\rowcolor{ideablue!8} Process & Population & Policy & Largest task & Task premium & Share of total & Minimum lower bound \\
\midrule
Bounded & ITT & Single & 125922 & 1.162 & 24.1\% & 0.077 \\
Bounded & ITT & Mean5 & 14969 & 3.458 & 50.9\% & 0.049 \\
Bounded & Faithful & Single & 14969 & 0.527 & 34.3\% & 0.008 \\
Bounded & Faithful & Mean5 & 14969 & 3.458 & 73.6\% & 0.013 \\
Agentic & ITT & Single & 9981 & 2.551 & 65.9\% & -0.014 \\
Agentic & ITT & Mean5 & 9981 & 5.710 & 70.8\% & 0.030 \\
Agentic & Faithful & Single & 9981 & 2.551 & 91.1\% & -0.118 \\
Agentic & Faithful & Mean5 & 9981 & 5.710 & 96.6\% & -0.079 \\
\bottomrule
\end{tabular}
\end{table}

\subsection{Development-to-fresh gain decomposition}
Let $g_D$ be the development-selected program's gain against the development baseline, $g_T$ its final gain against the final baseline, and $q_T$ the selected card's fresh final gain. Frozen weights give the identity
\[
g_D-q_T=(g_D-g_T)+(g_T-q_T).
\]
Each gain uses its split's baseline and the original operational utility, converted to accuracy-equivalent points. The first term combines selection optimism, development-to-final transfer, and the change from eight to nine training folds. The second is the prespecified premium. The full output provides all policies, groups, and both populations; the table gives pooled representative workflows.

\begin{table}[!ht]
\centering\small
\caption{Two additive sources of development-reported gain relative to fresh final gain. All intervals use paired task resampling with the robustness-analysis seed.}
\label{tab:gain-decomposition}
\begin{tabular}{llllll}
\toprule
\rowcolor{ideablue!8} Process & Population & Policy & Development--final & Final--fresh & Total \\
\midrule
Bounded & ITT & Single & 0.311 [-0.336, 0.981] & 0.268 [0.108, 0.447] & 0.580 [-0.098, 1.314] \\
Bounded & ITT & Mean5 & 0.236 [-0.599, 1.048] & 0.377 [0.081, 0.809] & 0.614 [-0.077, 1.317] \\
Bounded & Faithful & Single & 0.291 [-0.440, 1.094] & 0.096 [0.021, 0.185] & 0.387 [-0.351, 1.210] \\
Bounded & Faithful & Mean5 & 0.135 [-0.796, 1.066] & 0.294 [0.028, 0.748] & 0.429 [-0.278, 1.196] \\
Agentic & ITT & Single & 0.448 [-0.200, 1.205] & 0.215 [0.008, 0.533] & 0.663 [0.045, 1.379] \\
Agentic & ITT & Mean5 & 0.675 [-0.098, 1.472] & 0.448 [0.050, 1.125] & 1.123 [0.070, 2.338] \\
Agentic & Faithful & Single & 0.230 [-0.363, 0.966] & 0.156 [-0.081, 0.492] & 0.386 [-0.194, 1.077] \\
Agentic & Faithful & Mean5 & 0.372 [-0.341, 1.121] & 0.328 [-0.050, 0.994] & 0.700 [-0.292, 1.914] \\
\bottomrule
\end{tabular}
\end{table}

\subsection{Cost-ratio switches and mechanism families}
At a common continuous budget, the difference between the variance numerators for $K=2$ and $K=1$ is $\sigma_M^2c_R-\sigma_R^2c_M/2$. Thus the two-seed choice improves precision exactly when $c_M/c_R>2\sigma_M^2/\sigma_R^2$ for positive seed variance. Integer program counts can shift the finite-budget choice. The ratio applies to any consistent additive cost unit; monetary planning requires both model-request and evaluation prices.

Mechanism-family rows use complete primary model-judged faithful references, with at least two programs per card. The implementation estimate is pooled within-card seed-zero variance minus average within-program seed variance. Early-stopping raw estimates are negative ($-0.087$ and $-0.022$ squared points), so the working component is at its zero boundary. Its fitted threshold of zero reflects this boundary estimate. RBF programs show equal accuracies at all three recorded seeds; additional seeds supply no measured variance reduction in this bank. A zero observed component does not establish invariance over unobserved seeds.

\begin{table}[!ht]
\centering\small
\caption{Cost-ratio thresholds from pooled, source-stratum, and mechanism-family moments. Variances are in squared accuracy points. The observed pooled duration ratios are 2.56 and 3.72. Infinity denotes zero measured seed variance with a positive implementation component.}
\label{tab:cost-switches}
\begin{tabular}{llrrrc}
\toprule
\rowcolor{ideablue!8} Process & Population & $\widehat\sigma_M^2$ & $\widehat\sigma_R^2$ & $K=1\to2$ ratio & Equal-score programs \\
\midrule
Bounded & Pooled & 1.367 & 0.211 & 13.0 & -- \\
Bounded & Baseline HGB core & 1.649 & 0.280 & 11.8 & -- \\
Bounded & Early stopping & 0.000 & 0.841 & 0.0 & 45/207 \\
Bounded & Filtering & 0.646 & 0.004 & 290.9 & 200/212 \\
Bounded & Whitening & 4.484 & 0.004 & 2334.6 & 188/200 \\
Bounded & RBF & 0.572 & 0.000 & $\infty$ & 214/214 \\
Agentic & Pooled & 0.967 & 0.204 & 9.5 & -- \\
Agentic & Baseline HGB core & 1.092 & 0.260 & 8.4 & -- \\
Agentic & Early stopping & 0.000 & 0.834 & 0.0 & 45/202 \\
Agentic & Filtering & 0.925 & 0.005 & 383.0 & 202/213 \\
Agentic & Whitening & 2.265 & 0.003 & 1391.3 & 192/204 \\
Agentic & RBF & 0.705 & 0.000 & $\infty$ & 216/216 \\
\bottomrule
\end{tabular}
\end{table}

\subsection{A rule-selected follow-up example}
Eligible cards have twelve complete primary model-judged faithful reference programs and baseline values for all five statically checked gradient-boosting settings. We rank positive seed-zero implementation standard deviations by SD, task identifier, process, and card; for an even count we take the lower median. Among 68 eligible positive-SD cards, rank 34 is Bounded semeion I02, covariance whitening, with SD 0.487 accuracy points. The rule was specified before calculating the new paired comparisons, after the cohort outcomes were available.

Ten programs make 140 correct predictions and two make 138 on 160 shared items. Stable score ordering chooses reference slots 4 and 5 as the two central programs and slots 2 and 11 as the extreme pair. The central pair has identical full prediction vectors. Their observed difference and paired interval are zero, with exact McNemar $p=1$. For the extreme pair, two items favor the higher-scoring program and none favor the lower; the difference is 1.25 points, interval [0.00, 3.13], and exact $p=0.50$. The bootstrap conditions on these selected programs. Its interval for the extreme pair describes that realized contrast after outcome-based ordering. No significance threshold enters card selection or replacement.

The public supplement stores joint correctness-pattern counts for all twelve programs. Frozen learner settings are learning rate 0.08, 150 iterations, 31 leaves, minimum leaf size 20, and $L_2$ regularization 1. Whitening construction and numerical regularization vary within the card's permissions. The archived segment example and its paired counts remain a separate source illustration in Appendix~\ref{app:decision-diagnostics}.

\subsection{Targeted SVC convergence diagnostics}
Source analysis identifies seven literal positive iteration caps and one dynamic expression. We reconstruct these eight frozen programs, rerun the original final split at seed zero under the original scikit-learn 1.9.0 environment, and instrument the isolated worker to record fitted iteration counts, status, and warnings. Every rerun reproduces its saved prediction vector exactly. All seven literal caps reach 300 iterations in at least one binary subproblem and emit a convergence warning. The dynamic expression resolves to $-1$ on pc1 and converges in 1,103 iterations. Original scores and primary analyses are preserved.

\begin{table}[!ht]
\centering\small
\caption{Separate convergence reruns of the eight source-identified SVC candidates. All prediction vectors match the saved outcomes. The capped gesture program has 169/988 correct predictions and reaches its cap in all ten binary subproblems.}
\label{tab:svc-convergence}
\begin{tabular}{lllrrrrr}
\toprule
\rowcolor{ideablue!8} Cohort & Process & Task & Slot & Cap & Max iterations & At cap & Accuracy (\%) \\
\midrule
Archived & Bounded & mfeat-morphological & 1 & 300 & 300 & 1/45 & 70.00 \\
Archived & Agentic & phoneme & 0 & 300 & 300 & 1/1 & 69.69 \\
Initial & Bounded & GesturePhaseSegmentationProcessed & 6 & 300 & 300 & 10/10 & 17.11 \\
Followup & Agentic & pc1 & 3 & -1 & 1103 & 0/1 & 93.69 \\
Followup & Bounded & pc1 & 1 & 300 & 300 & 1/1 & 93.69 \\
Followup & Bounded & letter & 0 & 300 & 300 & 293/325 & 97.60 \\
Followup & Bounded & letter & 12 & 300 & 300 & 170/325 & 97.60 \\
Followup & Bounded & semeion & 8 & 300 & 300 & 9/45 & 93.75 \\
\bottomrule
\end{tabular}
\end{table}

\end{document}